\documentclass[longauth]{aa} 
\usepackage{amsmath} 
\usepackage{graphicx}
\usepackage{epstopdf}
\usepackage{natbib}
\bibpunct{(}{)}{;}{a}{}{,} 
\DeclareGraphicsExtensions{.jpg,.mps,.pdf,.png,.eps,.ps} 
\usepackage{txfonts}
\usepackage{url}
\usepackage{hyperref}
\usepackage{color}
\usepackage{longtable}
\usepackage{inputenc}
\begin{document}
\title{Physical properties of the trans-Neptunian object (38628) Huya from a multi-chord stellar occultation}


   \author{P. Santos-Sanz\inst{1}
		   \and	
		   J.L. Ortiz\inst{1}
		   \and		   
		   B. Sicardy\inst{2}
		   \and
		   M. Popescu\inst{3,4}
		   \and
		   G. Benedetti-Rossi\inst{2,5,6}
		   \and
		   N. Morales\inst{1}
		   \and
		   M. Vara-Lubiano\inst{1}
		   \and
		   J.I.B. Camargo\inst{7,5}
		   \and
		   C.L. Pereira\inst{7,5}
		   \and
		   F.L. Rommel\inst{7,5}
		   \and
		   M. Assafin\inst{8,5}
		   \and
		   J. Desmars\inst{9,10}
		   \and
		   F. Braga-Ribas\inst{11,2,5,7}
		   \and
		   R. Duffard\inst{1}
		   \and
		   J. Marques Oliveira\inst{2}
		   \and
		   R. Vieira-Martins\inst{5,7}
		   \and
		   E. Fern\'{a}ndez-Valenzuela\inst{12}
		   \and
		   B.E. Morgado\inst{2,5,7}
		   \and
		   M. Acar\inst{13,14}
		   \and
		   S. Anghel\inst{3,10,15}
		   \and
		   E. Atalay\inst{16}
		   \and
		   A. Ate\c{s}\inst{13}
		   \and
		   H. Bak{\i}\c{s}\inst{17}
		   \and
		   V. Bak{\i}\c{s}\inst{17}
		   \and
		   Z. Eker\inst{17}
		   \and
		   O. Erece\inst{17,18}
		   \and
		   S. Kaspi\inst{19}
		   \and
		   C. Kayhan\inst{13,20}
		   \and
		   S.E. Kilic\inst{18}
		   \and
		   Y. Kilic\inst{17,18}
		   \and
		   I. Manulis\inst{21}
		   \and
		   D.A. Nedelcu\inst{3}
		   \and
		   M.S. Niaei\inst{16}
		   \and
		   G. Nir\inst{21}
		   \and
		   E. Ofek\inst{21}
		   \and
		   T. Ozisik\inst{18}
		   \and
		   E. Petrescu\inst{22}
		   \and
		   O. Satir\inst{16}
		   \and
		   A. Solmaz\inst{23,24}
		   \and
		   A. Sonka\inst{3}
		   \and
		   M. Tekes\inst{23}
		   \and
		   O. Unsalan\inst{25}
		   \and
		   C. Yesilyaprak\inst{16,26}
		   \and
		   R. Anghel\inst{27}
		   \and
		   D. Berte\c{s}teanu\inst{28}
		   \and
		   L. Curelaru\inst{29}
		   \and
		   C. Danescu\inst{30}
		   \and
		   V. Dumitrescu\inst{31}
		   \and
		   R. Gherase\inst{28,32}
		   \and
		   L. Hudin\inst{33}
		   \and
		   A-M. Stoian\inst{34}
		   \and
		   J.O. Tercu\inst{34,35}
		   \and
		   R. Truta\inst{36}
		   \and
		   V. Turcu\inst{37}
		   \and
		   C. Vantdevara\inst{38}
		   \and
		   I. Belskaya\inst{39}
		   \and
		   T.O. Dementiev\inst{40}
		   \and
		   K. Gazeas\inst{41}
		   \and
		   S. Karampotsiou\inst{41}
		   \and
		   V. Kashuba\inst{42}
		   \and
		   Cs. Kiss\inst{43,44}
		   \and
		   N. Koshkin\inst{42}
		   \and
		   O.M. Kozhukhov\inst{40}
		   \and
		   Y. Krugly\inst{39}
		   \and
		   J. Lecacheux\inst{2}
		   \and
		   A. Pal\inst{43}
		   \and
		   \c{C}. P\"usk\"ull\"u\inst{45,46}
		   \and
		   R. Szakats\inst{43}
		   \and
		   V. Zhukov\inst{42}
		   \and
		   D. Bamberger\inst{47}
		   \and
		   B. Mondon\inst{48}
		   \and
		   C. Perell\'{o}\inst{49,50}
		   \and
		   A. Pratt\inst{51,50}
		   \and
		   C. Schnabel\inst{49,50}
		   \and
		   A. Selva\inst{49,50}
		   \and
		   J.P. Teng\inst{52}
		   \and
		   K. Tigani\inst{53}
		   \and
		   V. Tsamis\inst{53}
		   \and
		   C. Weber\inst{50}
		   \and
		   G. Wells\inst{47}
		   \and
		   S. Kalkan\inst{54}
           \and
   		   V. Kudak\inst{55}
           \and
           A. Marciniak\inst{56}
           \and
           W. Ogloza\inst{57}
           \and
           T. Ozdemir\inst{58}
           \and
           E. Pakstiene\inst{59}
           \and
           V. Perig\inst{55}
           \and
           M. Zejmo\inst{60}
		   }

\offprints{P. Santos-Sanz: psantos@iaa.es}

\institute{
Instituto de Astrof\'{\i}sica de Andaluc\'{\i}a (CSIC), Glorieta de la Astronom\'{\i}a s/n, 18008-Granada, Spain, 
\email{psantos@iaa.es}
\and
LESIA, Observatoire de Paris, PSL Research University, CNRS, Sorbonne Universit\'e, Univ. Paris Diderot, Sorbonne Paris Cit\'e, France
\and
Astronomical Institute of the Romanian Academy, 5 Cu\c{t}itul de Argint, 040557 Bucharest, Romania
\and
Instituto de Astrof\'{\i}sica de Canarias (IAC), C/V\'{\i}a L\'{a}ctea s/n, 38205 La Laguna, Tenerife, Spain
\and
Laborat\'{o}rio Interinstitucional de e-Astronomia -- LIneA \& INCT do e-Universo, Rua Gal. Jos\'{e} Cristino
77, Bairro Imperial de S\~ao Crist\'{o}v\~ao, Rio de Janeiro (RJ), Brazil
\and
UNESP - São Paulo State University, Grupo de Dinâmica Orbital e Planetologia, Guaratinguetá, SP, 12516-410, Brazil
\and
Observat\'{o}rio Nacional/MCTI, Rua Gal. Jos\'{e} Cristino 77, Bairro Imperial
de S\~ao Crist\'{o}v\~ao, Rio de Janeiro (RJ), Brazil
\and
Universidade Federal do Rio de Janeiro - Observat\'{o}rio do Valongo, Ladeira Pedro Antônio 43, CEP 20.080-090 Rio de Janeiro - RJ, Brazil
\and
Institut Polytechnique des Sciences Avancées IPSA, 63 boulevard de Brandebourg, F-94200 Ivry-sur-Seine, France
\and
Institut de M\'ecanique C\'eleste et de Calcul des \'Eph\'em\'erides, IMCCE, Observatoire de Paris, PLS Research University, CNRS, Sorbonne Universit\'es, UPMC Univ Paris 06, Univ. Lille, 77 Av. Denfert-Rochereau, F-75014 Paris, France
\and
Federal University of Technology--Paran\'{a} (UTFPR/Curitiba), Brazil
\and
Florida Space Institute, University of Central Florida, Orlando, USA
\and
ISTEK Belde Observatory, Turkey
\and
Department of Astronomy and Space Sciences, University of Ege, Turkey
\and
Faculty of Physics, University of Bucharest, 405, Atomistilor Street, 077125 Magurele, Ilfov, Romania
\and
Ataturk University, Astrophysics Research \& Application Center (ATASAM), 25240 Erzurum, Turkey
\and
Akdeniz University, Faculty of Sciences, Department of Space Sciences and Technologies, 07058 Antalya, Turkey
\and
T\"UBITAK National Observatory, Akdeniz University Campus, 07058 Antalya, Turkey 
\and
School of Physics and Astronomy and Wise Observatory, Tel Aviv University, Israel
\and
Department of Astronomy and Space Sciences, Science Faculty, Erciyes University, 38030 Melikgazi, Kayseri, Turkey
\and
Particle Physics and Astrophysics Department, Weizmann Institute of Science, Israel
\and
Astronomical Observatory ``Amiral Vasile Urseanu'' Bucharest, Romania
\and
Space Sciences and Solar Energy Research and Application Center (UZAYMER), \c{C}ukurova University, Adana, Turkey
\and
Space Observation and Research Center, \c{C}ağ University, Mersin, Turkey
\and
Ege University, Faculty of Science, Department of Physics, 35100, Bornova, Izmir, Turkey
\and
Ataturk University, Science Faculty, Department of Astronomy \& Space Sciences, 25240 Erzurum, Turkey
\and
Astronomical Observatory ``Victor Anestin'' Bacau, Romania
\and
Astroclubul Bucure\c{s}ti, Romania
\and
L13 Observatory, Romania
\and
L15 Observatory, Romania
\and
Ia cu Stele, Romania
\and
L16 Observatory, Romania
\and
L04 Observatory, Romania
\and
Galaţi Astronomical Observatory, ``Răsvan Angheluţă" Museum Complex of Natural Sciences, 6A, Regiment 11 Siret Street, 800340, Galaţi, Romania
\and
Tiraspol State University, 5, Ghenadie Iablocikin Street, MD-2069, Chișinău, Republic of Moldova
\and
Asociatia Astroclubul Quasar, Romania
\newpage
\and
Romanian Academy, Cluj-Napoca Branch, Astronomical Observatory Cluj-Napoca, Str. Cireşilor, Nr. 19, 400487, Cluj-Napoca, Romania
\and
L22 Observatory, Romania
\and
Institute of Astronomy, V.N. Karazin Kharkiv National University, Ukraine
\and
QOS Observatory, National Space Facilities Control and Test Center, Ukraine
\and
Section of Astrophysics, Astronomy and Mechanics, Department of  Physics, National and Kapodistrian University of Athens, GR-15784  Zografos, Athens, Greece
\and
Astronomical Observatory of Odessa I.I. Mechnikov National University, Shevchenko Park, Odessa, 65014, Ukraine
\and
Konkoly Observatory, Research Centre for Astronomy and Earth Sciences, E\"otv\"os Lor\'and Research Network (ELKH), H-1121 Budapest, Konkoly Thege Miklós út 15-17, Hungary
\and
ELTE E\"otv\"os Lor\'and University, Institute of Physics, Budapest, Hungary
\and
Çanakkale Onsekiz Mart University, Faculty of Arts and Sciences, Department of Physics, TR-17020, Çanakkale, Turkey
\and
Çanakkale Onsekiz Mart University, Astrophysics Research Center and Ulupınar Observatory, TR-17020, Çanakkale, Turkey
\and
Northolt Branch Observatories, UK
\and
Sainte Marie, La Reunion, France
\and
Agrupaci\'{o} Astron\`{o}mica de Sabadell, Barcelona, Spain
\and
International Occultation Timing Association--European Section (IOTA-ES), Germany
\and
Almalex Observatory, Leeds, UK
\and
Les Makes, La Reunion, France
\and
Ellinogermaniki Agogi Observatory, Athens, Greece
\and
Ondokuz Mayis University Observatory, Space Research Center, 55200, Kurupelit, Samsun, Turkey
\and
Laboratory of Space Researches, Uzhhorod National University, Uzhhorod, Daleka Str., 2A, 88000, Ukraine
\and
Astronomical Observatory Institute, Faculty of Physics, A. Mickiewicz University, Słoneczna 36, 60-286 Poznań, Poland
\and
Mt. Suhora Observatory, Pedagogical University, ul. Podchorazych 2, PL-30-084 Krakow, Poland
\and
Department of Physics, Faculty of Science and Arts, İnönü University, Malatya, Turkey
\and
Astronomical Observatory, Institute of Theoretical Physics and Astronomy, Vilnius University, Sauletekio av. 3, 10257 Vilnius, Lithuania
\and
Janusz Gil Institute of Astronomy, University of Zielona Góra, Prof. Szafrana 2, PL-65-516 Zielona Góra, Poland
}

   \date{Received  / Accepted }


 
  \abstract
   {Within our international program to obtain accurate physical properties of trans-Neptunian objects (TNOs) we predicted a stellar occultation by the TNO (38628) Huya of the star \textit{Gaia} DR2 4352760586390566400 (m$\rm_G$ =  11.5 mag.) for March 18, 2019. After an extensive observational campaign to obtain astrometric data we updated the prediction and it turned out to be favorable to central Europe. Therefore, we mobilized half a hundred of professional and amateur astronomers in this region and the occultation was finally detected from 21 telescopes located at 18 sites in Europe and Asia. This makes the Huya event one of the best ever observed stellar occultation by a TNO in terms of the number of chords.}
   {The aim of our work is to determine accurate size, shape and geometric albedo of the TNO (38628) Huya by using the observations obtained from a multi-chord stellar occultation. We also aim at providing constraints on the density and other internal properties of this TNO.}
   {The 21 positive detections of the occultation by Huya allowed us to obtain well-separated chords which permitted us to fit an ellipse for the limb of the body at the moment of the occultation (i.e., the instantaneous limb) with kilometric accuracy.}
   {The projected semi-major and minor axes of the best ellipse fit obtained using the occultation data are (a$^\prime$, b$^\prime$) = (217.6 $\pm$ 3.5 km , 194.1 $\pm$ 6.1 km) with a position angle of the minor axis P$^\prime$ = 55.2$^\circ \pm$ 9.1. From this fit, the projected area-equivalent diameter is of 411.0 $\pm$ 7.3 km. This diameter is compatible with the equivalent diameter for Huya obtained from radiometric techniques (D = 406 $\pm$ 16 km). From this instantaneous limb, we obtained the geometric albedo for Huya ($p_V = 0.079 \pm 0.004$) and we explored possible three-dimensional shapes and constraints to the mass density for this TNO. We did not detect the satellite of Huya through this occultation, but the presence of rings or debris around Huya is constrained using the occultation data. We also derived an upper limit for a putative Pluto-like global atmosphere of about $p_{\rm surf}$ = 10 nbar.}
   {    }
   
\titlerunning{Physical properties of (38628) Huya from a stellar occultation}
   \maketitle
%

\section{Introduction}
The stellar occultation technique is a very direct way to obtain highly accurate sizes, derive albedos, and, in some cases, even densities and 3D shapes for TNOs \citep[e.g.][]{Sicardy2011,Ortiz2012,Ortiz2017}. Atmospheres and satellites can also be detected and characterized by means of stellar occultations \citep{Sicardy2006,Meza2019}. Very fine details, undetectable by any other ground-based technique, can also be detected using stellar occultations, such as the rings detected around the centaurs Chariklo \citep{Braga-Ribas2014} and Chiron \citep{Ortiz2015,Ruprecht2015} and around the dwarf planet Haumea \citep{Ortiz2017}. These discoveries have opened a new way of research within the planetary sciences of the distant solar system bodies \citep{Sicardy2019a,Sicardy2019b}. All of the above points out that the stellar occultation technique is a very powerful means to obtain information about the physical properties of TNOs. 

From October 2009, when the first stellar occultation by a TNO, apart from Pluto, was recorded \citep{Elliot2010}, to date, about 77 stellar occultations produced by 33 different TNOs, excluding the one presented here, have been detected. About 50 of these occultations have been detected from only one or two different locations which did not allow us to obtain the projected shape and size of the body. In these cases, it is always possible to derive astrometric positions with very low uncertainties for the TNO that can be used to improve their ephemerides \citep{Rommel2020}. To be able to fit the five parameters of an ellipse, at least three different observations are needed, otherwise, the problem is degenerate. The remaining 22 occultations (produced by 15 different TNOs) of the total of 77, were detected from more than two locations (i.e., multi-chord occultations) which allowed us to properly characterize their shapes and to derive very interesting and accurate physical properties. This means that we have only obtained a good physical characterization using stellar occultations of about 15 TNOs, excluding Huya \citep[see e.g.][and references therein]{Elliot2010,Sicardy2011,Ortiz2012,Ortiz2017,Ortiz2020b,Braga-Ribas2013,Benedetti-Rossi2016,Benedetti-Rossi2019,Dias-Oliveira2017,Souami2020}.  

The stellar occultation by the TNO (38628) Huya presented here is the best ever stellar occultation by a TNO in terms of the number of chords obtained so far, excluding Pluto and the recent occultation by the TNO (307261) 2002 MS$_4$ on August 8, 2020 (Rommel et al. in prep). It is also the first multi-chord occultation reported for Huya.

The TNO (38628) Huya was discovered on March 10, 2000 from the `Llano del Hato National Astronomical Observatory' in M\'erida (Venezuela) by astronomers of the Quasar Equatorial Survey Team \citep[QUEST, ][]{Ferrin2001}. Huya is classified as a plutino (i.e. it is in the 2:3 mean motion orbital resonance, MMR, with Neptune) and it is among the group of the $\sim$ 100 largest known TNOs, in fact, it is one of the biggest plutinos together with (119951) 2002 KX$_{14}$, (84922) 2003 VS$_2$, and (208996) 2003 AZ$_{84}$. 

Infrared spectra of Huya appears moderately red and featureless with a lack of signatures of water ice and other volatiles \citep{Jewitt2001,deBergh2004,Fornasier2013}, however other infrared spectra show faint signs of water ice \citep{Barucci2011,Merlin2017}. This points out to a surface covered with a thick and red layer of dark organic compounds which would be homogeneously covered with trace amounts of water ice. Aqueously altered silicate minerals have also been proposed to explain some absorption features observed in the near-infrared spectra of Huya \citep{Licandro2001,deBergh2004}. Besides spectra, colors from Spitzer Space Telescope at 3.6 and 4.5 $\mu$m, have been modeled indicating a surface composition of 40 $\pm$ 20\% H$_2$O, 30 $\pm$ 10\% silicates, and 30 $\pm$ 10\% organics \citep{Fernandez-Valenzuela2021}.

The rotational period of Huya is not well-determined due to the small peak-to-peak amplitude of its light curve, but a period of 6.75 $\pm$ 0.01 h was proposed by \citet{Ortiz2003} based on observations taken in February and March 2002. According to more recent work by \citet{Thirouin2014}, the period of 6.75 h could be an alias of Huya's rotation period that they propose (5.28 h) based on observations obtained from 2010 to 2013. Other alternative periods cannot be totally ruled out. 

Huya has a known satellite provisionally designated as S/2012 (38628) 1, discovered by a team led by K. Noll in 2012 using images from Hubble Space Telescope (HST) \citep{Noll2012}. The satellite's separation distance from the primary is estimated to be at least 1740 km. 

The thermal emission from the Huya's system has been measured with Spitzer/MIPS in 2004 and with Herschel/PACS and SPIRE in 2010. From these thermal data, using H$_{V}$ = 5.04 $\pm$ 0.03 mag, \cite{Fornasier2013} obtained an area-equivalent diameter for the system of D$\rm_{eq} = 458 \pm 9$ km and a geometric albedo at V-band of p$\rm_V = 0.083 \pm 0.004$. The area-equivalent diameter of Huya itself (D$\rm_{Huya}$ = 406 $\pm$ 16 km) and its satellite (D$\rm_{Satellite} \sim$ 213 km), assuming the same albedo as Huya, were also obtained in that work. A summary of the orbital elements and most relevant physical characteristics of Huya from the literature is shown in Table \ref{OrbPhys_Huya}.

In the present work, we analyzed the 21 chords obtained from the stellar occultation of the star \textit{Gaia} DR2 4352760586390566400 (m$\rm_G$ =  11.5 mag.) produced on March 18, 2019 by the TNO Huya. This paper is organized as follows. In Section \ref{observations} we describe the observations carried out to predict the occultation and the observations of the occultation itself. In Section \ref{analysis} we detail the analysis of the Huya's occultation data. In Section \ref{rotLC} the observations performed to refine the rotational period of Huya in order to determine the rotational phase at the moment of the occultation are presented. In Section \ref{results} the results obtained from the analysis of the occultation are given and discussed (i.e. the instantaneous limb fit, the projected diameter, the 3D shapes, the albedo, the density and the search for material orbiting Huya, etc). Lastly, our conclusions are presented in Section \ref{conclu}.

\begin{table*}
\caption{Orbital and physical parameters of Huya from literature.} 
\label{OrbPhys_Huya} 
\centering 
\scalebox{0.82}{%
\begin{tabular}{l c c c c l c c c c c c c} 
\hline\hline 
Object	&	a 	&	q 	&	i 	&	e	&	H$_{V}$	&	P	&	$\Delta$m	&	Class. & D$\rm_{Huya}$ & p$\rm_{V}$ & D$\rm_{Satellite}$\\
	&	[AU]	&	[AU]	&	[deg]	&		&[mag]& 	[h]	&	[mag]	&  & [km] & [\%] & \\
\hline 
(38628) Huya	&	39.69 &	28.55	&	15.47 	&	0.28	&	5.04 $\pm$ 0.03$^{a}$ & 6.75 $\pm$ 0.01$^{b}$ / 5.28 $\pm$ 0.01$^{c}$	&	$<$ 0.1$^{b}$ / 0.02 $\pm$ 0.01$^{c}$ &	Plutino	&	406 $\pm$ 16$^{a}$ & 0.083 $\pm$ 0.004$^{a}$ & $\sim$ 213 km$^{a}$\\
\hline 
\end{tabular}}

\small
{\bf a:} semi-major axis in Astronomical Units (AU), {\bf q:} perihelion distance in AU, {\bf i:} orbital inclination in degrees, and {\bf e:} eccentricity, from Minor Planet Center (MPC-IAU) database, May 2021. {\bf H$_V$ [mag]:} absolute magnitude at V-band. {\bf P [h]:} preferred single-peaked rotational periods. {\bf $\Delta$m [mag]:} peak-to-peak amplitude of the rotational light curves shown in the previous column. {\bf Class.}: dynamical classification following \cite{Gladman2008} scheme. {\bf D$\rm_{Huya}$:} area-equivalent diameter of Huya from radiometric technique, {\bf p$_{\rm V}$:} geometric albedo at V-band from radiometric technique, {\bf D$\rm_{Satellite}$:} estimated diameter of the Huya's satellite from radiometric technique. {\bf References:} a) \cite{Fornasier2013}; b) \cite{Ortiz2003}; c) \cite{Thirouin2014}.


\end{table*}



\section{Observations}
\label{observations}


\subsection{Predictions}
\label{pred}

Due to the wealth of information that can be obtained from stellar occultations by TNOs \citep[e.g.][and references therein]{Sicardy2011,Ortiz2012,Ortiz2017,Ortiz2020a,Ortiz2020b}, we have been performing intensive astrometric and photometric observing campaigns since 2010 with the aim to predict these events. This observing strategy allows us to derive accurate predictions \citep[see][for a recent review]{Ortiz2020a}.

Huya's orbit and ephemerides were derived using the Numerical Integration of the Motion of an Asteroid, NIMA \citep{Desmars2015}, a tool developed within the ERC Lucky Star project\footnote{https://lesia.obspm.fr/lucky-star/}. Part of the astrometric data of Huya used to feed NIMA were obtained with the 1.6-m telescope at Observat\'{o}rio Pico dos Dias (OPD) in Brazil on June 3, 2017 and July 8, 17, 2018. This telescope was equipped with an Andor IKon-L camera (2048 $\times$ 2048 pixels, field of view -FOV- = 6.1 $\times$ 6.1 arcmin, Resolution = 0.18 arcsec/pixel) and I-Johnson filter. In the course of these campaigns we found that the TNO (38628) Huya would occult a V = 11.7 mag star (\textit{Gaia} DR2 4352760586390566400, m$\rm_G$ =  11.5 mag) on March 18, 2019.

Around five months before the event, we refined the original NIMA's prediction using more astrometric data, as well as data available in the Minor Planet Center database\footnote{https://www.minorplanetcenter.net/db-search}. Near the occultation date, we refined again the prediction using images of Huya acquired with the 1.5-m telescope in Sierra Nevada Observatory (OSN) in Granada, Spain. This observing run was performed on March 2, 3, 8, 10, and 11, 2019, around two weeks before the occultation event, with the 2k $\times$ 2k Andor IKON-L camera. 109 images were obtained in 2 $\times$ 2 binning mode at moon illumination $< 20\%$. The detailed setup of these observations, including weather conditions and other related information, is shown in Table \ref{ObsSummary}. Bias and sky flat-field frames were taken each night to calibrate the images.

\begin{table*}
	\centering
	\caption{Summary of the `last minute' astrometric observing campaign.}
	\label{ObsSummary}
	\resizebox{\textwidth}{!}{%
	\begin{tabular}{cccccccccccccc}   
		\hline	    
\textbf{Telescope} & \textbf{Date} & \textbf{CCD} & \textbf{Scale} & \textbf{FOV} & \textbf{Filter} & \textbf{Exp.} & \textbf{N} & \textbf{Seeing} & \textbf{SNR} & \textbf{1$\sigma$(RA)} & \textbf{1$\sigma$(Dec)} & \textbf{Offset(RA)} & \textbf{Offset(Dec)} \\
\hline
\textbf{OSN 1.5-m} & March 2,3,8,10,11, 2019 & 2k $\times$ 2k & 0.46$''$/pix & 7.5$'$ $\times$ 7.5$'$  & R-Johnson & 400 s & 109 & 1.6'' & 45 & 8 mas & 8 mas & 149 mas & -92 mas \\

\hline
	\end{tabular}}
      \small
   	\textbf{Telescope} is the telescope used during the observing run. \textbf{Date} are the observing dates. \textbf{CCD} is the size of the Andor IKON-L detector. \textbf{Scale} is the binned image scale of the instrument. \textbf{FOV} is the field of view of the instrument. \textbf{Filter} is the filter used. \textbf{Exp.} is the exposure time of the individual images. \textbf{N} is the total number of images. \textbf{Seeing} is the average seeing during the observing runs. \textbf{SNR} is the signal to noise ratio of the object. \textbf{1$\sigma$(RA)} is the average 1$\sigma$ uncertainty in right ascension (RA) of the astrometry in milliarcseconds (mas). \textbf{1$\sigma$(Dec)} is the average 1$\sigma$ uncertainty in declination (Dec) of the astrometry in mas. The \textbf{Offset(RA)} and \textbf{Offset(DEC)} are the offsets for right ascension and declination with respect to the JPL\#28 orbit expressed in mas.

\end{table*}

\begin{figure}[!hpbt]
\centering
    \includegraphics[width=\columnwidth]{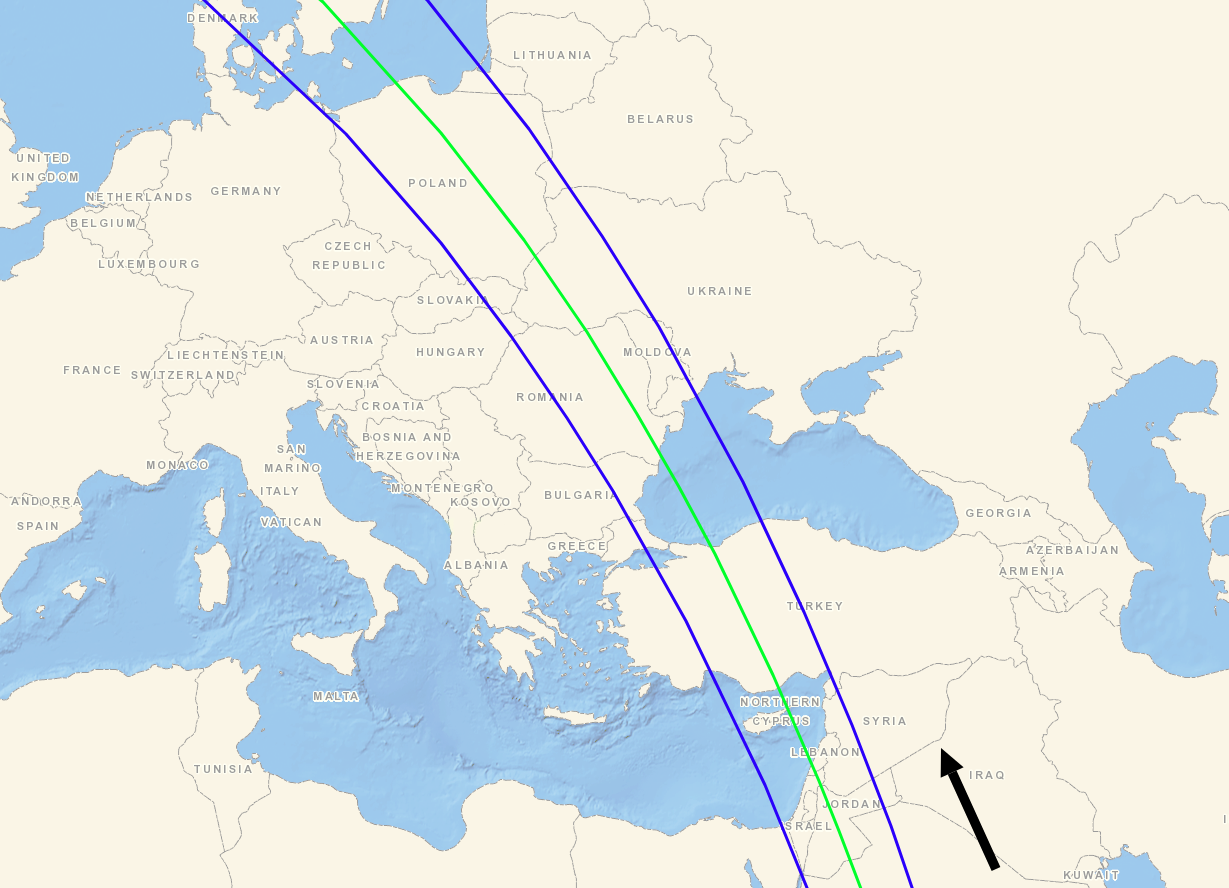}
    \caption{Last minute prediction of the occultation using {\it Gaia} DR2 star catalog \citep{Gaia2016,Gaia2018} and relative astrometry of Huya regarding the occulted star obtained from the 1.5-m telescope in Sierra Nevada Observatory, Granada, Spain. The green line indicates the centrality of the shadow path and the blue lines the limits of the shadow assuming an area-equivalent diameter for Huya of 406 km \citep{Fornasier2013}. The 1$\sigma$ precision along the path is 165 km and the 1$\sigma$ precision in time is 20 s. The arrow in the right bottom shows the direction of the shadow motion. Map credit: \url{https://www.gpsvisualizer.com/} and Australian Topography (\textsuperscript{\textcopyright}Commonwealth of Australia --Geoscience Australia-- 2016. Creative Commons Attribution 4.0 International License).}
    \label{predictmap} 
\end{figure}

The images obtained during the OSN observing campaign were astrometrically solved using \textit{Gaia} DR2 \citep{Gaia2016,Gaia2018}, which was the most precise astrometric catalog available when this occultation happened. The prediction was obtained using the coordinates of the occulted star (\textit{Gaia} DR2 4352760586390566400) propagated to the occultation epoch using the \textit{Gaia} DR2 proper motions and parallax, and the relative astrometry (offsets) of Huya with respect to the Jet Propulsion Laboratory orbit JPL\#28. The average 1$\sigma$ astrometric uncertainties obtained and the offsets with respect to the orbit JPL\#28 are shown in Table \ref{ObsSummary}. These uncertainties translate to 1$\sigma$ uncertainties of 165 km projected on Earth surface (cross-track)
and 20 s in time (along-track). The final prediction map obtained is shown in Figure \ref{predictmap}.  Note that the center line of the shadow path of this final prediction is $\sim$ 3000 km east of the prediction based only on the JPL\#28 orbit of Huya and the \textit{Gaia} DR2 position of the star, and $\sim$ 1900 km west of the prediction obtained about five months before the event. The width of the shadow path in this Figure is assumed to be the diameter of Huya derived from Herschel and Spitzer thermal data \citep[D = 406 km, according to][]{Fornasier2013}. The real shadow path of the occultation was $\sim$ 78 km west of the final prediction (see Figure \ref{shadowpath}) but within the estimated uncertainties.

As the `last minute' prediction from the OSN telescope (see Table \ref{ObsSummary}) was favorable to many countries in Europe and involved a bright star, we alerted our (previously notified) European collaborators emphasizing the importance of observing this occultation. 21 positive detections from 18 sites longitudinally distributed and 16 misses were obtained.

\subsection{Stellar Occultation}
\label{OccObs}

On March 18, 2019, a total of 49 telescopes located in Europe and Asia (Israel) were ready to observe the occultation by the TNO Huya. Some of them were located far from the nominal prediction (e.g. Spain, UK, Eastern Turkey) to allow the detection of the Huya satellite, possible rings, and also because we alerted the community of amateur occultation observers through the `planoccult' list. Time series of images or video observations were acquired with 37 of these telescopes (from the other 12, 8 had bad weather, and 4 suffered technical problems). Video observations were converted to FITS images prior to their analysis using same procedures and precautions adopted in \cite{Benedetti-Rossi2016}. From 21 of the 37 telescopes, located at 18 different sites, we recorded the occultation of the star produced by Huya. From 16 of the 37 telescopes the occultation was not detected. All stations and telescopes that recorded a positive detection plus two very close stations which missed the occultation (QOS Observatory in Ukraine and \c{C}ukurova University Observatory in Turkey) are listed in Table \ref{ObservDetails}. Basic information about the 49 telescopes that participated in this campaign and their locations is shown in the online Table \ref{SummaryObserv}.

Aperture photometry for each data set was obtained, from which we derived the light curves (flux of the star normalized to the mean value before the occultation, versus time). The light curves from the 21 positive observations at 18 sites (see Table \ref{ObservDetails}) showed drops in flux caused by the occultation. Light curves from the other 16 telescopes that did not detect the occultation were also obtained and carefully analyzed. We analyzed with special care the negative detection at QOS Observatory in Zalistci (Ukraine), which was the closest to the shadow path. This closer non detection can notably constrain the projected shape edge of Huya (see Figure \ref{shadowpath} and Table \ref{ObservDetails}).

\begin{figure}[!hpbt]
\centering
	\includegraphics[width=\columnwidth]{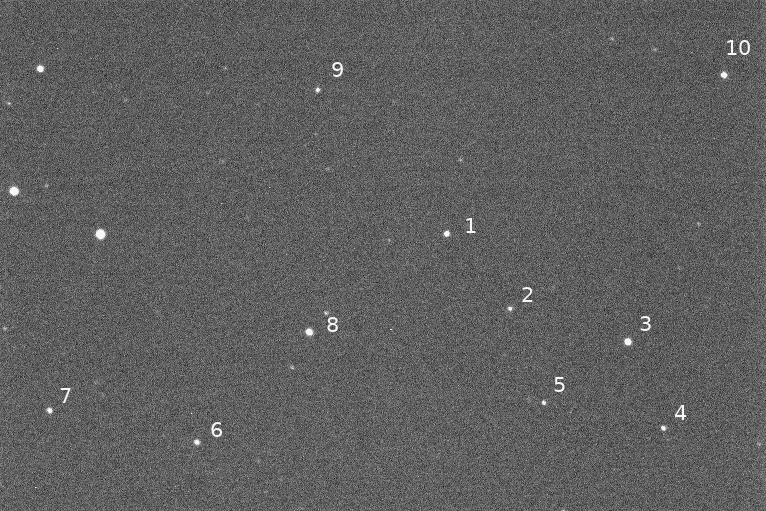}
    \caption{One of the images of the occultation obtained from the Astronomical Observatory Cluj-Napoca in Romania (see Table \ref{ObservDetails} for details). The source marked with 1 corresponds to the occulted star plus Huya before the beginning of the occultation. The other marked sources were used as reference stars. The FOV of the image is 20 $\times$ 13  arcmin, north is up and east is to the left.
    }
    \label{fov_example}
\end{figure}

The time in all the observing sites was synchronized via Network Time Protocol servers (NTP) or GPS-based Video Time inserters (VTI). Each image header includes the acquisition time. Most of the series of images were obtained with around 10-15 minutes before the predicted occultation time up to around 10-15 minutes after the event. This strategy allows us to have a good characterization of the photometric baseline and to probe the surroundings of the object (i.e. the presence or not of satellites, rings or other orbiting material). No filters were used in any of the telescopes in order to maximize the signal to noise ratio (SNR) of the occulted star with the aim to obtain the best possible photometric data.

The light curves of the occultation were obtained using our own IDL (Interactive Data Language) codes based on DAOPHOT routines and using the PRAIA photometry tool \citep{Assafin2011}, after the usual dark, bias, and flat-field corrections when calibration images were available. Then, the relative photometry of the occulted star was obtained on the images using the stars present in the FOV as comparison stars with the aim of minimizing flux variations due to atmospheric transparency fluctuations and to different seeing conditions. We used as many reference stars as possible from 3 to $\sim$ 25 depending on the FOV (see Fig. \ref{fov_example}). In some cases, the observers only read a Region of Interest (RoI) of the detector to minimize the readout time, in these cases, the FOV was smaller (i.e., a less number of reference stars) than the FOV from other telescopes. We did not use the same reference stars for different data sets, as the number of stars within the FOV was different from telescope to telescope. In the end, the flux of the occulted star is compared to an ensemble mean of all the selected reference stars. The chosen aperture diameters, starting with diameters around the full width at half maximum (FWHM), minimized the flux dispersion of the occulted star before and after the occultation.

It is important to note that the flux of the occulted star obtained during the occultation is the combination of the flux of the star and the flux of Huya, although in this case, the flux of Huya has a very small contribution ($\sim$ 0.07 \%) to the total flux due to the brightness of the occulted star. Finally, the light curve (i.e. the combined flux of the occulted star and Huya versus time) was obtained for each data set. The light curves obtained from the 21 telescopes that detected the occultation show deep dimmings in flux at the expected occultation times (Figure \ref{OccLCs}). Detailed information about the sites, telescopes, detectors, exposure times, observers, and light curves dispersion, from which the occultation was detected is shown in Table \ref{ObservDetails}. This Table also includes information about the two stations closest to the shadow path that reported a negative detection (QOS Observatory in Ukraine and \c{C}ukurova University Observatory in Turkey). More details and complementary information on the analysis of these occultation light curves are given in Section \ref{analysis}.

\begin{figure}[!hpbt]
	\includegraphics[width=\columnwidth]{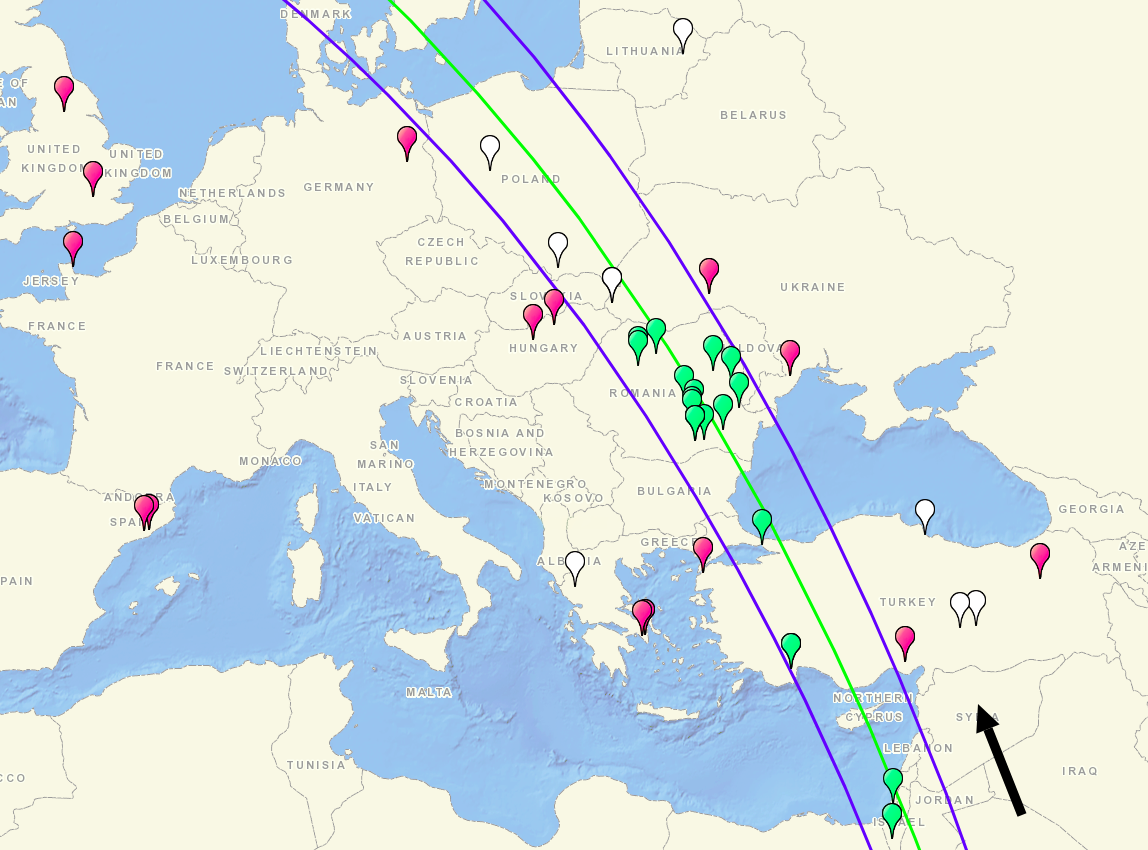}
    \caption{Map of post-occultation shadow path of the stellar occultation by Huya, the path width (limited by the blue lines) is the area-equivalent diameter obtained from the reconstructed ellipse fit (D = 411.3 km, see Section \ref{shape-albedo}).  The site positions from where the occultation was recorded are shown with green marks, the negative detections with red marks, and the sites that did not observe due to bad weather or to technical problems are indicated with white marks. The green line indicates the centrality of the shadow path and the blue lines the limits of the shadow. The arrow in the right bottom shows the direction of the shadow motion. Map credit: \url{https://www.gpsvisualizer.com/} and Australian Topography (\textsuperscript{\textcopyright}Commonwealth of Australia --Geoscience Australia-- 2016. Creative Commons Attribution 4.0 International License).
    }
    \label{shadowpath}
\end{figure}


\section{Analysis of the Stellar Occultation by Huya}
\label{analysis}

The TNO (38628) Huya occulted the m$\rm_{G}$= 11.5 mag star {\it Gaia} DR2 4352760586390566400 on March 18, 2019 at $\sim$ 00:53 UTC. The typical sampling times were greater than 0.2 seconds (i.e., $\ge$ 1.61 km in the body plane of Huya), this means that our data are dominated by the exposure times rather than by the Fresnel diffraction effects ($\simeq$ 1.13 km) or the stellar diameter ($\simeq$ 0.70 km at Huya's distance). The Fresnel scale value of F = $\sqrt{\lambda d/2}$ = 1.13 km is obtained from the geocentric distance of Huya during the occultation (d = 28.3500 AU) and the average central wavelength of the observations ($\lambda$ = 600 nm). To estimate the angular diameter of the occulted star we use its V (11.751 mag) and K (10.294 mag) apparent magnitudes from the NOMAD catalog \citep{Zacharias2004} and the \cite{vanBelle99} equation for main sequence stars, obtaining a diameter of $\simeq$ 0.0342 mas ($\simeq$ 0.70 km at Huya's distance). Table \ref{StarDetails} summarizes the occulted star details and other related information.

\begin{table*}
	\centering
	\caption{Details of the occulted star and other occultation related information.}
	\label{StarDetails}
	\resizebox{\textwidth}{!}{%
	\begin{tabular}{l|l}   
		\hline	    
\textbf{Designation}   &  {\it Gaia} DR2 4352760586390566400  \\
\hline
\textbf{Coordinates DR2}$^{a}$  & 	$\alpha$ = 16h 41m 06s.4260, $\delta$ = -06$^\circ$ 43' 34''.5756\\ 
\hline
\textbf{Star position errors}$^{b}$ & errRA = 0.1474 mas, errDec = 0.1143 mas \\ 
\hline
\textbf{Proper motions \& Parallax}$^{c}$ & pmRA = -4.506 $\pm$ 0.090 mas yr$^{-1}$, pmDec = -14.340 $\pm$ 0.055 mas yr$^{-1}$, Plx = 1.5717 $\pm$ 0.0478 mas \\ 
\hline
\textbf{Magnitudes}$^{d}$    & 	B = 12.296, V = 11.751, R = 11.380, J = 10.541, H = 10.384, K = 10.294, G= 11.538 \\
\hline
\textbf{Star Diameter}$^{e}$    & 	$\sim$ 0.0342 $\pm$ 0.0014 mas ($\sim$ 0.70 $\pm$ 0.03 km at Huya's distance) \\	
\hline
\textbf{Fresnel}$^{f}$    & 	    1.13 km  \\	
\hline
\textbf{Exposure time effect}$^{g}$    & 	1.61 km    \\	
\hline
\textbf{Velocity}$^{h}$  & 	8.07 km/s	\\
		\hline
	\end{tabular}}
\begin{list}
      \small
   	  \item $^{a}${\it Gaia} DR2 coordinates propagated to the occultation epoch (2019.211) using the proper motions and parallax.
      \item $^{b}$ errRA and errDEc are the errors in RA and Dec from the {\it Gaia} DR2 catalog propagated to the occultation epoch using the formalism by \cite{Butkevich2014}. This formalism has been applied using the SORA tool \citep{Gomes-Junior2022}
      \item $^{c}$ pmRA and pmDec are the proper motions in RA and Dec and respective errors, Plx is the absolute stellar parallax with error.
      \item $^{d}$B, V, R, J, H and K from the NOMAD catalog \citep{Zacharias2004}. G from the {\it Gaia} DR2 catalog \citep{Gaia2018}.
      \item $^{e}$Size estimated using V and K magnitudes and the \cite{vanBelle99} equation for a main sequence star.
	  \item $^{f}$Fresnel diffraction effect: F = $\sqrt{\lambda d/2}$, with d = 28.3500 AU at the occultation moment and $\lambda$ = 600 nm.
	  \item $^{g}$Smallest exposure time from the positive detections multiplied by the event shadow velocity.
	  \item $^{h}$Velocity of Huya with respect to the star as seen from Earth.
\end{list}
\end{table*}

\begin{figure*}[!hpbt]
	\includegraphics[width=\textwidth]{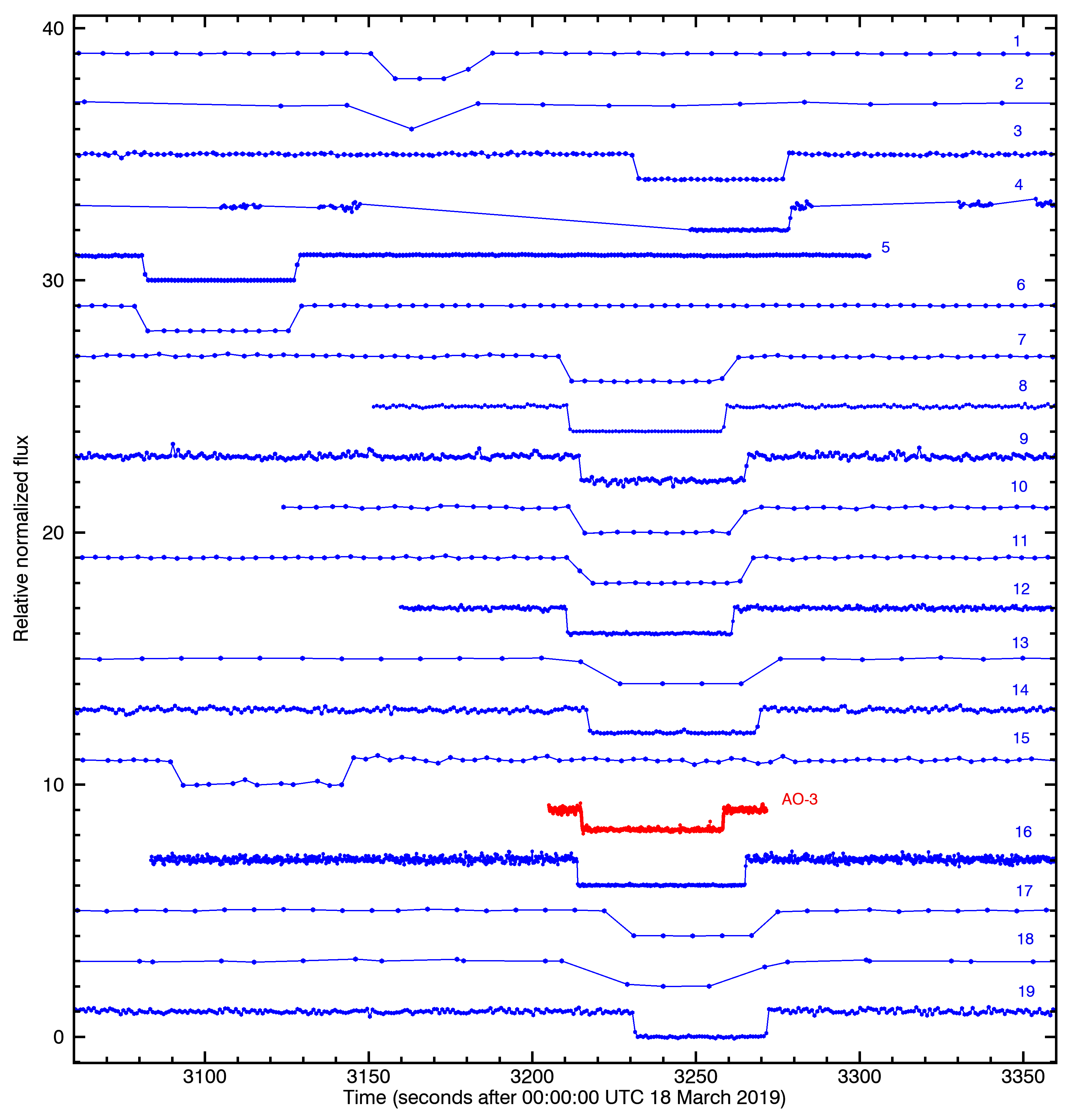}
    \caption{Stellar occultation light curves (normalized flux versus time) from the 21 positive detections, as presented in Table \ref{ObservDetails}. Each individual measurement is indicated with a dot, these measurements have been connected with a solid line for clarity. The light curves are shifted in flux for a better viewing and are presented from top to bottom with respect to their distance to the center of the predicted shadow path, West to East. Flux uncertainties are not shown in order to avoid an unreadable plot (the standard deviation of the measurements are indicated in Table \ref{ObservDetails}). The chord \#18 is the combination of 2 light curves obtained from the same site (Galați Observatory, Romania) with 2 telescopes of 40-cm and 20-cm, respectively. The light curve in red (AO-3) was not used to obtain the limb fit, it has been shifted in time to properly appear in this plot because the time information of this chord was missed.}
    \label{OccLCs}
\end{figure*}

The light curves obtained from the stellar occultation (see Figure \ref{OccLCs}) are used to derive the times of disappearance (`ingress' time) and reappearance (`egress' time) of the star behind the Huya's limb. These ingress and egress times can be directly translated to distances in the plane of the sky using the apparent motion of Huya relative to the occulted star, which is 8.07 km/s for this occultation (see Table \ref{StarDetails}). The derived segments in the plane of the sky are known as `chords'. From these chords we can obtain the physical properties described in Section \ref{results}. The times of disappearance and reappearance and their uncertainties at each site are obtained creating a synthetic light curve by using a square-well model convolved with the Fresnel scale, the star apparent size and the exposure time \citep{Elliot1984,Roques1987}. The parameters fitted are the ingress and egress times and the depth of the occultation. The difference between the data and the synthetic light curve is iteratively minimized using a $\chi^2$ metric. This technique is described in detail in \cite{Ortiz2017} and \cite{Benedetti-Rossi2019} and references therein. 

The ingress and egress times and their associated uncertainties obtained in this way are shown in Table \ref{Chords}. These times determined 20 chords of different sizes in the plane of the sky also included in Table \ref{Chords}. Due to a CCD camera failure the ROASTERR-1 Observatory (Romania) missed the ingress. As a consequence only the egress time is considered for this observatory (see light curve \#4 in Figure \ref{OccLCs}). At the end, 19 chords obtained from the 21 telescopes that detected the occultation in Table \ref{ObservDetails} were used to obtain the results described in Section \ref{results}:

\begin{itemize}

\item The chord from the Amateur Observatory-3 in Romania (AO-3) is not used  because the absolute time information of this chord was missed, there was no timestamp in images and no time in image headers, this means that we only have relative timing. We used reference time for the first frame as 00:00:00.0 UTC and derive the chord size with its associated error bars.

\item The light curves obtained from Galați Observatory-1 and Galați Observatory-2 in Romania were obtained at the same site with the same integration time, so we merged the two data sets in order to have a single light curve and chord (\#18) with more data with the star occulted.

\end{itemize}

\section{Rotational light curve of Huya}
\label{LC_obs}

To obtain the rotational phase of Huya at the moment of the occultation, we used the already published photometry of 299 images acquired in 2010-2013 with the 1.5-m OSN telescope in Granada (Spain) and with the 1.23-m telescope at Calar Alto Observatory in Almer\'ia (Spain), see \cite{Thirouin2014} for further details about these observations. We merged these data with 116 new images of Huya acquired on 8 nights on July 1-4 and August 1-4 ($\sim$ 3-4 h of observation each night), 2019 with the 1.5-m OSN telescope with integration times of 400 seconds, binning 2 $\times$ 2 and no filter in order to achieve the highest SNR. 

Standard bias and flat field corrections were applied on all the images before the extraction of the fluxes of the object and of the selected comparison stars by means of aperture photometry technique. To perform this task we used specific routines coded in IDL, trying different values for the apertures (starting with apertures around the FWHM) and sky ring annulus for the object and the same values for the comparison stars in order to maximize the SNR of the object and to minimize the dispersion of the photometry. When possible, we choose the same set of comparison stars for all nights within the observing run to minimize systematic photometric errors. The data processing used was the same as that described in detail in e.g. \cite{Fernandez-Valenzuela2016}. 

\begin{figure}[!ht]
	\includegraphics[width=\columnwidth]{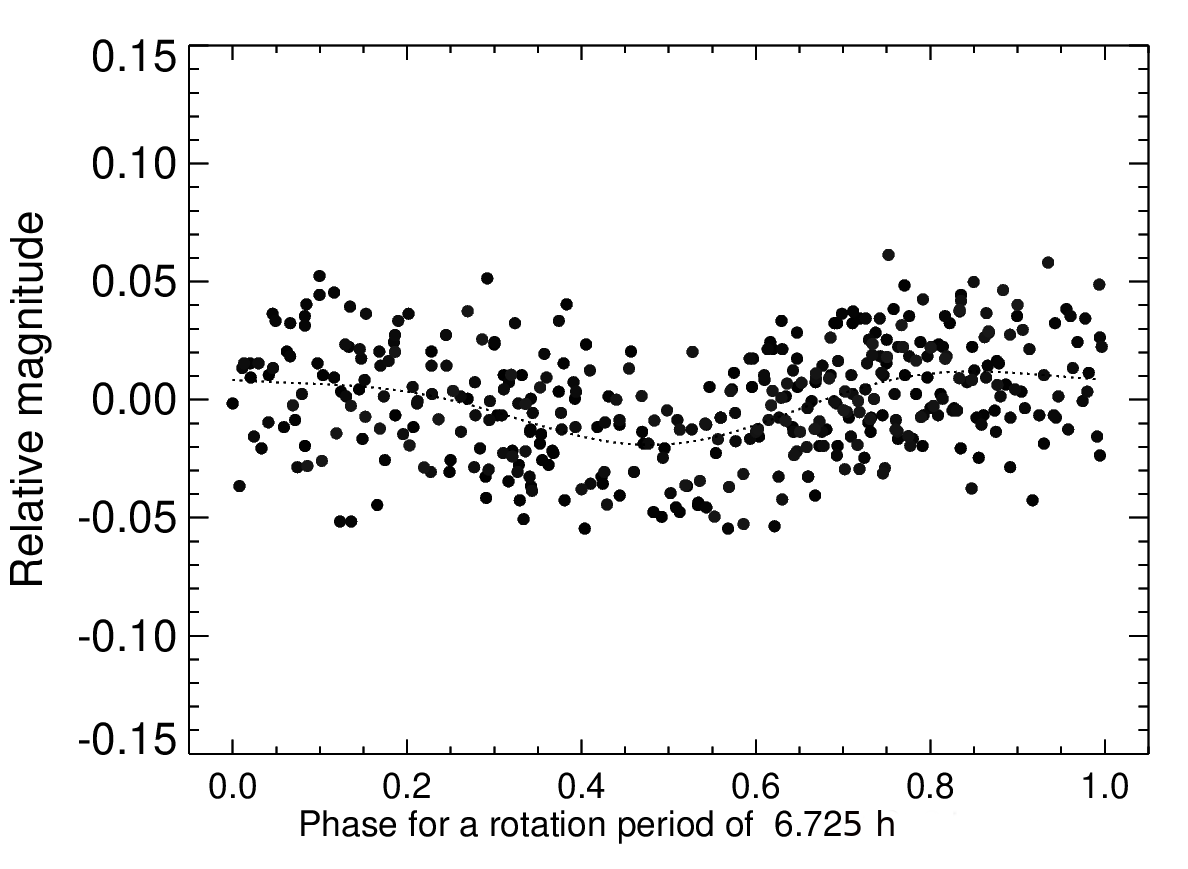}
	\includegraphics[width=\columnwidth]{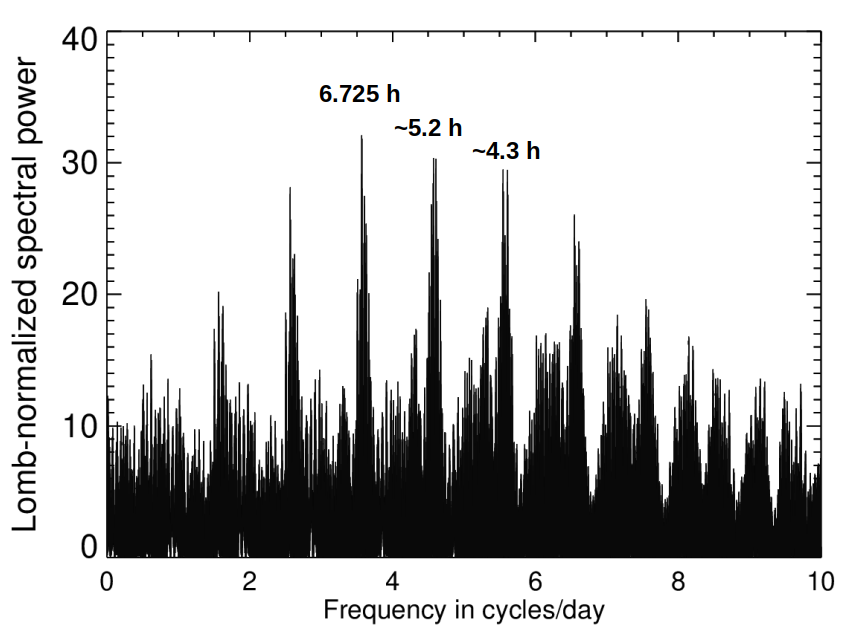}
    \caption{{\bf Top panel:} Rotational light curve (relative magnitude vs. rotational phase) of Huya obtained from the data described in Section \ref{LC_obs}. The data have been folded with a rotational period of 6.725 h. The zero rotational phase in the plot is fixed close to the moment of the occultation (March 18, 2019 00:54:00 UTC). {\bf Bottom panel:} Lomb periodogram showing the rotational period P = 6.725 h, which corresponds to the peak with the highest spectral power. The two other aliases at high spectral power correspond to other possible rotational periods of Huya: P $\sim$ 5.2 h and P $\sim$ 4.3 h.}
    \label{rotLC}
\end{figure}

The final product obtained is the flux of Huya respect to the comparison stars versus time (corrected for light travel times). A Lomb-Scargle technique \citep{Lomb1976} is applied to these time series data in order to obtain the rotational period of Huya. The rotation period with the highest spectral power obtained from the Lomb-Scargle periodogram is 6.725 $\pm$ 0.006 h (Figure \ref{rotLC}), which is compatible with other published rotation periods \citep[e.g.][]{Ortiz2003,Thirouin2014}. Aliases of this period, like e.g. $\sim$ 5.2 h or $\sim$ 4.3 h, cannot be excluded as is shown in the bottom panel of Figure \ref{rotLC}.

This updated rotational light curve allows us to determine the rotational phase at the moment of the occultation and it turns out that Huya was near one of its absolute brightness minima at the time of the March 18, 2019 occultation. This means that the TNO occulted the star when its apparent surface area was near its minimum. The rotational light curve in the upper panel of Figure \ref{rotLC} is fitted with a Fourier function in order to derive the peak-to-peak amplitude of the light curve, which results to be of 0.031 $\pm$ 0.005 mag. This amplitude is slightly larger than the previously published amplitudes \citep{Ortiz2003,Thirouin2014}. The zero rotational phase in the upper panel of Figure \ref{rotLC} was chosen to be the moment of the stellar occultation (i.e. March 18, 2019 00:54:00 UTC).

Alternatively, the very shallow rotational light curve obtained could be directly related to the rotation of the satellite of Huya, as it is suspected to be the case for the TNO 2002 TC$_{302}$ \citep{Ortiz2020b}. If Huya's satellite has a very irregular shape, it could cause a shape-driven light curve which would dominate the rotational light curve of the system, this would also be consistent with a very round shape for Huya (Maclaurin spheroid). On the other hand, a rotational period of 6.725 h would indeed be a weird period for a satellite, because a rotation period synchronized with its orbital period is expected, which would be of days, not of hours. However, we know that Haumea's largest satellite, Hi'iaka, has a much faster rotation period ($\sim$ 9.8 h) than its orbital period ($\sim$ 49.5 days) according to \cite{Hastings2016}. A binary system with a mass ratio of the satellite to the main body similar to the Huya system is Varda-Ilmarë (mass ratio $\sim$ 9\%). In this system, it is likely that the rotation of the main body, Varda, is not synchronized with the mutual orbital period. This can be explained by the fact that the estimated time scale for synchronization is longer than the age of the solar system \citep{Grundy2015}. Similar calculations can be performed when the orbit of the Huya’s satellite becomes available, which will shed light on whether or not its rotation is synchronized with its orbital period, and whether this potential lack of synchronization is common in TNO satellites.

\section{Results}
\label{results}

The obtained physical properties of Huya, namely, its projected shape, the size, the geometric albedo, and the 3D shape are shown in sections \ref{shape-albedo} and \ref{3D-models}. These were determined from the chords presented in the previous section. In Section \ref{rings} the environment around Huya is characterized and possible rings or debris orbiting this plutino are constrained. Finally, in Section \ref{atmosphere} we constrain a putative Huya's atmosphere.


\subsection{Projected shape, diameter and albedo of Huya}
\label{shape-albedo}

From the 19 chords numbered in Table \ref{Chords}, we have 37 chord extremities that can be used to obtain the projected shape (i.e. the instantaneous limb) of Huya at the moment of the occultation. These extremities give the position ($f$, $g$) of the star projected in the sky plane of the body and relative to the center of the object. $f$ and $g$ are counted positively toward local celestial east and celestial north, respectively, and are measured in kilometers \citep[see, e.g.,][]{Benedetti-Rossi2019}. Note that we have 37 instead of 38 extremities because we discard the ingress time of the ROASTERR-1 Observatory in Romania (chord \#4), which was missed due to a CCD camera failure. The closest negative chord from QOS Observatory in Ukraine is also used to constrain the limb. We fitted an ellipse to these 37 points as described below. 

\begin{figure}[!hpbt]
	\includegraphics[width=\columnwidth]{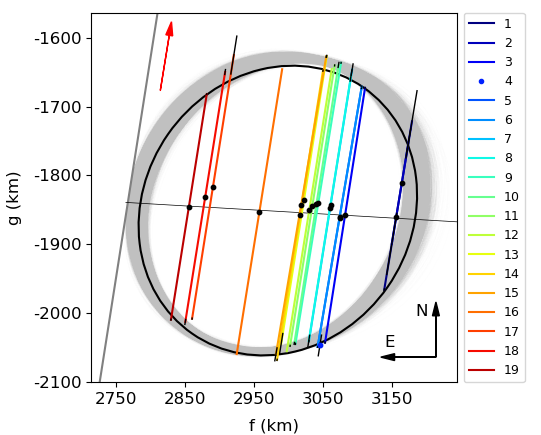}
    \caption{Best-fitting ellipse (in black) to the 19 chords presented in Table \ref{Chords} (for chord \#4 only egress time is used). The fitted ellipse determines the projected shape of Huya at the moment of the occultation and has axes of 435.2 $\pm$ 7.0 km $\times$ 388.2 $\pm$ 12.2 km. The gray ellipses are the best-fitting ellipses to the $10^4$ randomly generated chord extremities (see Section \ref{shape-albedo} for details). The black dots are the centers of the chords, and the gray line crossing all the chords is a weighted linear fit to these dots. The black solid lines in the extremities of the chords are the 1$\sigma$ uncertainties of the ingress/egress times. The red arrow shows the direction of the shadow motion. The chord numbers are the same as those used in Tables \ref{ObservDetails} and \ref{Chords} and Figure \ref{OccLCs}. The negative detection at QOS Observatory in Ukraine (light gray line at the left of the ellipse) helps to constrain the limb fit.}
    \label{elliptical_fit}
\end{figure}

The best-fitting ellipse was obtained through a $\chi^2$ minimization method applied to the 37 chord extremities. The function to minimize is $\chi^2 = \sum_{i=1}^{37}(r_{i,\rm obs}-r_{i,\rm com})^2 / \sigma^{2}_{i,\rm r}$, where r is the radius from the center of the ellipse ($f_{\rm c}$ , $g_{\rm c}$), the subscripts ``obs'' and ``com'' means observed and computed respectively and $\sigma_{i,\rm r}$ are the errors on the extremities determination. What we finally obtain from this fit are the apparent or projected semi-axes of the ellipse (a$^\prime$ , b$^\prime$), the center of the ellipse ($f_{\rm c}$ , $g_{\rm c}$) and the position angle of the minor axis of the ellipse (P$^\prime$), i.e. the apparent position angle of the pole measured eastward from celestial north. Since the ellipse fit is not linear in the parameters, an estimate of the error in the parameters using a $\chi^2$ minimization technique may result in an underestimation of the errors. Therefore, we estimated the uncertainties in the five parameters of the fitted ellipse using a Monte Carlo approach, as done in \cite{Santos-Sanz2021}. We randomly generated the 37 chord extremities $10^4$ times, meeting the requirements provided by their corresponding uncertainties. For each one of these sets of randomly generated extremities, we obtained the best-fitting ellipse in terms of a minimization of the sum of squared residuals $\sum_{i=1}^{37}(r_{i,\rm obs}-r_{i,\rm com})^2$, where $r_{i,\rm obs}$ and $r_{i,\rm com}$ are the same as described above. In the end, we had $10^4$ best-fitting ellipses to the 37 randomly obtained extremities (i.e. $10^4$ possible values of the five parameters that determine an ellipse: a$^\prime$, b$^\prime$, $f_{\rm c}$, $g_{\rm c}$, and P$^\prime$). From these parameter distributions, we obtained the 1$\sigma$ uncertainty in each parameter as the standard deviation of the corresponding distribution (see Table \ref{limbfit} and Figure \ref{elliptical_fit}).

The parameters and uncertainties obtained for the best-fitting ellipse as explained above were ($f_{\rm c}$ , $g_{\rm c}$) = (2984.7 $\pm$ 3.2 km , -1850.9 $\pm$ 1.7 km), (a$^\prime$ , b$^\prime$) = (217.6 $\pm$ 3.5 km , 194.1 $\pm$ 6.1 km) and P$^\prime$ = 55.2$^\circ \pm$ 9.1. The axes ratio of this ellipse is small and close to 1.0 (a$^\prime$/b$^\prime$ = 1.12 $\pm$ 0.05) which indicates that Huya is a very round object, as is shown in Figure \ref{elliptical_fit} where the 19 chords (only egress for chord \#4) and their uncertainties obtained from the occultation are plotted in the plane of the sky together with the best-fitting ellipse to their extremities. Note that the center obtained from the elliptical fit ($f_{\rm c}$ , $g_{\rm c}$) provides the offsets with respect to the positions of Huya obtained using the Jet Propulsion Laboratory orbit JPL\#28 + Development Ephemeris model DE431, assuming that the {\it Gaia} DR2 occulted star position is correct. 
As shown in Figure \ref{elliptical_fit}, the centers of the chords were not well aligned and their extremities were not always well fitted by the best-fitting ellipse. On the other hand, an ellipse would be the expected projected shape for Huya considering that an icy body with its size should have adopted a regular equilibrium figure with only minor irregularities due to topography (see discussion in Section \ref{3D-models}). If this is true, the chords centers must be aligned \citep[see, e.g.][]{Braga-Ribas2013,Santos-Sanz2021}. For completeness, and to check if this solution is very different from the original one we slightly shift the chords by performing a linear fit to the centers of the chords and looking for the ellipse that best fits these chords. The semi-axes obtained for the best-fitting ellipse to the shifted chords were (a$^\prime$ , b$^\prime$) = (220.2 $\pm$ 4.3 km , 197.4 $\pm$ 6.1 km), which differ by less than 2\% concerning the unshifted solution. Since both solutions, the original and the shifted one, are virtually the same, we decided to use the best-fitting ellipse to the original (unshifted) chords as the best solution since in the latter no a priori assumption is made about the centers of the chords.

From the limb fit to the original chords we derived an area-equivalent diameter of Huya at the moment of the occultation of $D_{eq}$ = 411.0 $\pm$ 7.3 km. If Huya is a triaxial body this diameter is a lower limit of the total area-equivalent diameter of this plutino because, as shown in Section \ref{LC_obs}, Huya was very close to its absolute brightness minimum at the moment of the occultation, i.e. its projected area was very close to its minimum. If Huya is an oblate Maclaurin-like body this diameter will be the real projected diameter, and the rotational light curve should be due, in this case, to albedo variations in the surface. Anyway, this diameter is smaller than the radiometric area-equivalent diameter obtained using Herschel (PACS and SPIRE) and Spitzer (MIPS) measurements: $D_{eq}$ = 458 $\pm$ 9.2 km \citep{Fornasier2013}. 

However, to do a proper comparison of the occultation and radiometric diameters we have to take into account that the radiometric diameter includes the thermal flux of Huya and its satellite since neither Herschel Space Observatory nor Spitzer Space Telescope can resolve the Huya's system. \cite{Fornasier2013} did an estimation of the equivalent size of the main body and the satellite from radiometric models (assuming same geometric albedo for both bodies) obtaining $D_{\rm Main}$ = 406 $\pm$ 16 km and $D_{\rm Satellite}$ = 213 $\pm$ 30 km. This means that the area-equivalent diameter of Huya obtained from the occultation is slightly bigger (411.0 km) than the radiometric-derived one (406 km), but both diameters are fully compatible within their error bars. This underestimation in the radiometric diameters has been noted for other TNOs when properly comparing occultation diameters with radiometric ones \citep[][]{Ortiz2020b,Ortiz2020a}.

From the best-fitting ellipse we derived the geometric albedo at V-band of the Huya's surface at the moment of the occultation by means of the equation \citep{Sicardy2011}: 

\begin{equation}
p_{V} = 10^{0.4(V_{sun}-H_{\rm V})}/(A/\pi)
\end{equation}

, where $V_{sun}$ is the V-magnitude of the Sun ($V_{sun}$ = -26.74 mag), $H_{\rm V}$ is the absolute magnitude of Huya at V-band, and $A$ is the projected area of the TNO (expressed in AU$^2$) directly obtained from the occultation limb fit.

It is important to highlight that the absolute magnitude provided for Huya in the literature is that of the Huya's system (i.e., the combined magnitude of Huya and its satellite). To obtain the absolute magnitude of Huya itself we used the difference in optical magnitude between Huya and its satellite measured by \cite{Noll2012} which is of $\sim$ 1.4 mag. Using this difference, and assuming $H_{\rm V}$ = 5.04 $\pm$ 0.03 mag for the Huya's system from \cite{Fornasier2013}, we obtained an absolute magnitude of $H_{\rm V}$ = 5.31 mag for Huya. Finally, to obtain the albedo, we have to correct this $H_{\rm V}$ taking into account the rotational phase at the moment of the occultation. The object was close to its minimum projected area during the occultation (Section \ref{LC_obs}) this means that we should add $\Delta m/2$ = 0.031/2 mag = 0.016 mag to $H_{\rm V}$ to derive the absolute magnitude of Huya at the moment of the occultation, obtaining $H_{\rm V}$ = 5.326 mag. Using the latter $H_{\rm V}$ we obtained a geometric albedo of $p_{\rm V}$ = 0.079 $\pm$ 0.004 for Huya. This albedo is smaller than the albedo derived from the radiometric method using Herschel and Spitzer thermal data \citep[$p_{\rm V} = 0.083 \pm 0.004$, from][]{Fornasier2013}, but the radiometric albedo was computed for the Huya's system.


\subsection{Three-dimensional shape models for Huya}
\label{3D-models}

In what follows, we have assumed fluid-like behavior for Huya, as is commonly done for TNOs \citep[see, e.g.,][]{Thirouin2010,Duffard2009}, and have used the Chandrasekhar formalism \citep{Chandrasekhar1987} to obtain possible Maclaurin or Jacobi equilibrium shapes for this TNO.

In general, TNOs with shallow rotational light curves ($\Delta$m $\le$ 0.15 mag) are associated with Maclaurin-like shapes \citep{Sheppard2002b,Ortiz2003,Duffard2009}. In this case, the variability is produced by albedo marks on the surface of the object, and the rotational light curves will be single-peaked. For larger peak-to-peak amplitudes ($\Delta$m $>$ 0.15 mag) a Jacobi ellipsoid shape is expected. In this case, the rotational light curve is shape-driven, presenting a double-peaked shape. We believe that most ($\sim$ 75\%) of the large TNOs (D $>$ 300 km) have Maclaurin spheroid shapes, rather than triaxial ellipsoid shapes \citep{Duffard2009}.

According to \cite{Tancredi2008}, Huya has enough size to reach hydrostatic equilibrium. We also know from the statistics of rotational periods and amplitudes of TNOs, most of which are the size of Huya or even smaller, that hydrostatic equilibrium shapes reproduce that statistic well \citep[see, e.g,][]{Duffard2009}. This is a clear indication that not only Huya but also other smaller TNOs seem to be compatible with hydrostatic equilibrium. One of the reasons could be that the mechanical properties of the internal material are probably weaker than those of water ice, perhaps due to porosity or other effects \citep[see, e.g,][]{Thirouin2010}. Then, under this equilibrium assumption, a Maclaurin spheroid with semi-axes a, b, and c (a = b $>$ c, where c is the rotation axis) is the most likely three-dimensional shape for this TNO, taking into account the single-peaked and shallow ($\Delta$m = 0.031 mag) rotational light curve presented by this body (see Section \ref{LC_obs} and Figure \ref{rotLC}). Therefore, we explore possible Maclaurin shapes compatible with the rotational light curve and with the results obtained from the occultation. According to \cite{Bierson2019} and from Figure 3 in \cite{Grundy2019} the expected density for a TNO with D $\sim$ 410 km is $\rho$ $<$ 800 kg/m$^3$. A Maclaurin spheroid with this density rotating with a period of 6.725 h (i.e., the preferred one discussed in Section \ref{LC_obs}) would have semi-axis ratios a/c = b/c = 1.87. The aspect angle\footnote{We consider that the aspect angle is 0$^\circ$ if the object is in a pole-on geometry and 90$^\circ$ for an equator-on geometry.} needed to obtain, from this Maclaurin, the projected axial ratio obtained from the occultation (a$^\prime$/b$^\prime$ = 1.12) is 32$^\circ$, which is a very small aspect angle. Note that for a random distribution of spin axes, the most likely aspect angle is 60$^\circ$ and the typical probability density distribution goes with the sine of the aspect angle. The likelihood of aspect angles 32$^\circ$ or smaller is $\sim$ 15\%. This probability is not too small, but it means it is not very likely.

If Huya's satellite is in the equatorial plane, the tilt angle of the orbit will allow us to determine the aspect angle. Huya's satellite orbit is not available yet although it should be ready soon (Grundy private communication). However, with the data we currently have, a Maclaurin spheroid  with  $\rho \sim$ 800 kg/m$^3$ compatible with the occultation and with the preferred rotational period (P = 6.725 h) would have axes: 2a = 435 km, 2b = 435 km, 2c = 233 km, for an aspect angle of 32$^\circ$. For the other possible and shorter rotation periods (P $\sim$ 5.2 h and P $\sim$ 4.3 h) the aspect angle should be even smaller ($< 30^\circ$), which is more unlikely (probability $\sim$ 13\%).

Alternatively, we can assume that the rotational light curve of Huya is produced by a rotating triaxial ellipsoid (i.e. a Jacobi ellipsoid) with semi-axes a, b and c (a $>$ b $>$ c, with c the axis of rotation). The minimum density expected for a Jacobi ellipsoid rotating with P = 6.725 h is $\rho$ = 859 kg/m$^3$. The aspect angles compatible with the occultation (a$^\prime$/b$^\prime$ = 1.12) and with the rotational light curve ($\Delta$m $<$ 0.031 mag) span from 31$^\circ$ to 43$^\circ$. For these aspect angles, a family of Jacobi solutions rotating at 6.725 h with $\rho_{\rm min}$ = 859 kg/m$^3$ is possible: 2a = [440-455] km, 2b = [428-442] km, 2c = [253-261] km. On the other hand, the minimum density required for a Jacobi rotating with P $\sim$ 5.2 h is $\rho$ = 1440 kg/m$^3$, a very large density for a body of the Huya's size. A Jacobi ellipsoid rotating with P $\sim$ 4.3 h would require an even larger and unlikely minimum density ($\rho$ = 2100 kg/m$^3$).

An alternative three-dimensional shape solution to a Maclaurin (close to pole-on) or Jacobi (with too high densities) shapes invokes non-hydrostatic equilibrium shapes, via granular media or differentiation (the latter unlikely for Huya's size) which would not need so high densities. The exploration of these non-hydrostatic solutions is out of the scope of this paper.


\subsection{Probing Huya's environment}
\label{rings}

We did not detect secondary drops below the 3$\sigma$ level of noise in the positive occultation light curves related to the satellite or possible rings around Huya (Figure \ref{OccLCs}). Secondary drops below the 3$\sigma$ level were also not detected in the negative light curves. We can constrain the presence of rings (or debris) around Huya using the light curves obtained during the occultation \citep[e.g.][]{Braga-Ribas2014,Ortiz2017,Sickafoose2019,Santos-Sanz2021}. 

For a given exposure time in seconds (t$_{\rm exp}$), and a measured photometric uncertainty ($\sigma$), for event velocity v (v = 8.07 km/s for the occultation by Huya), at a significance of 3$\sigma$, the minimum detectable ring width is w = t$_{\rm exp}$ v. Opacity variations can be detected at the 3$\sigma$ level, where op = 3$\sigma$. For other opacities (op) the minimum width (w) of a ring detectable at 3$\sigma$ can be obtained as w = 3$\sigma$ t$_{\rm exp}$ v / op.

The best positive occultation light curve, in terms of flux dispersion, was obtained from the C18 telescope at the Wise Observatory data (see Table \ref{ObservDetails}) with $\sigma_{flux}$ = 0.009. The minimum opacity of a putative ring detectable at 3$\sigma$ using this light curve is 2.7\% for a ring width of 24.2 km. In other words, a ring with a width $\geq$ 24.2 km and an opacity $\geq$ 2.7\% would produce a drop in the light curve easily detectable below the 3$\sigma$ level of noise. If the opacity of the ring were 50\% we could have detected rings with a width $\geq$ 1.3 km at 3$\sigma$. For an opacity of 100\% a ring with a width $\geq$ 0.7 km could have been detected below the 3$\sigma$ level (see Table \ref{ringconstraints}). The latter constraints on the ring's width were obtained without take into account the dead-time during the data acquisition, which was of 1.46 s for the C18 at Wise Observatory. This means that rings with widths $\leq$ 11.8 km could have been lost during the dead-times. The other positive occultation light curve obtained at the same observatory with the W-FAST telescope \citep{Nir2021} has a very short dead-time of only 0.007 s (maximum ring width during dead-time = 0.06 km) with also a small dispersion in flux ($\sigma_{flux}$ = 0.015). Therefore, from this data we could have detected rings at the 3$\sigma$ level of noise with widths $\geq$ 0.7 km for an opacity of 50\% and with widths $\geq$ 0.4 km for an opacity of 100\%, in both cases, with negligible ring width losses during the dead-times ($\leq$ 0.06 km).

We obtained constraints at the 3$\sigma$ level on the ring sizes for the other positive light curves and the closest negative one for the minimum opacity and for opacities of 50\% and 100\%. The derived constraints are shown in Table \ref{ringconstraints}. The best constraint, in terms of ring width, was obtained from the Amateur Observatory-3 in Romania, with integration times of 0.0333 s (without dead-times) and $\sigma_{flux}$ = 0.074, from where rings with widths $\geq$ 0.1 km for an opacity of 50\% could have been detected at the 3$\sigma$ level of noise.

From all the above, we can conclude that no debris or rings of the type found in the dwarf planet Haumea \citep[opacity $\sim$ 50\%,][]{Ortiz2017} have been detected around Huya through this stellar occultation. However, narrow ($\leq$ 0.1 km) and optically thin rings (opacity $\le$ 50\%) at different geometries, not probed by the occultation light curves, cannot be ruled out. Note also that diffraction effects will be important for rings with a width $<$ 1.13 km (the Fresnel scale value) and high opacity.


\subsection{Constraints to a putative atmosphere}
\label{atmosphere}

Assuming that Huya is a predominantly icy body, we attempt to place constraints on a possible atmosphere. The sublimation rate of water ice is extremely low at Huya's distance from the Sun. Only very volatile ices can sublimate there, but volatile retention models such as those by \cite{Schaller2007} indicate that a body of Huya's size cannot have retained those ices. So the most likely scenario is that Huya does not have an atmosphere of any sort. The ingress and egress data can shed light on an upper limit for a potential atmosphere, but not a very tight one. In any case, we estimate here the upper limit of an atmosphere around Huya, and more precisely, its surface pressure $p_{\rm surf}$. To do so, we have to assume a composition and a temperature profile for this putative atmosphere, which are both unknown.

We may derive some orders of magnitude by assuming that Huya's atmosphere has the same composition (mainly N$_2$ with traces of CH$_4$), and same temperature profile as Pluto's atmosphere. This results in a surface temperature of about 36 K, with a rapid increase up to $\sim$110 K at 30 km altitude due to CH$_4$ heating, followed by a roughly isothermal upper branch near 100 K.

\begin{figure}[!hpbt]
	\includegraphics[width=\columnwidth]{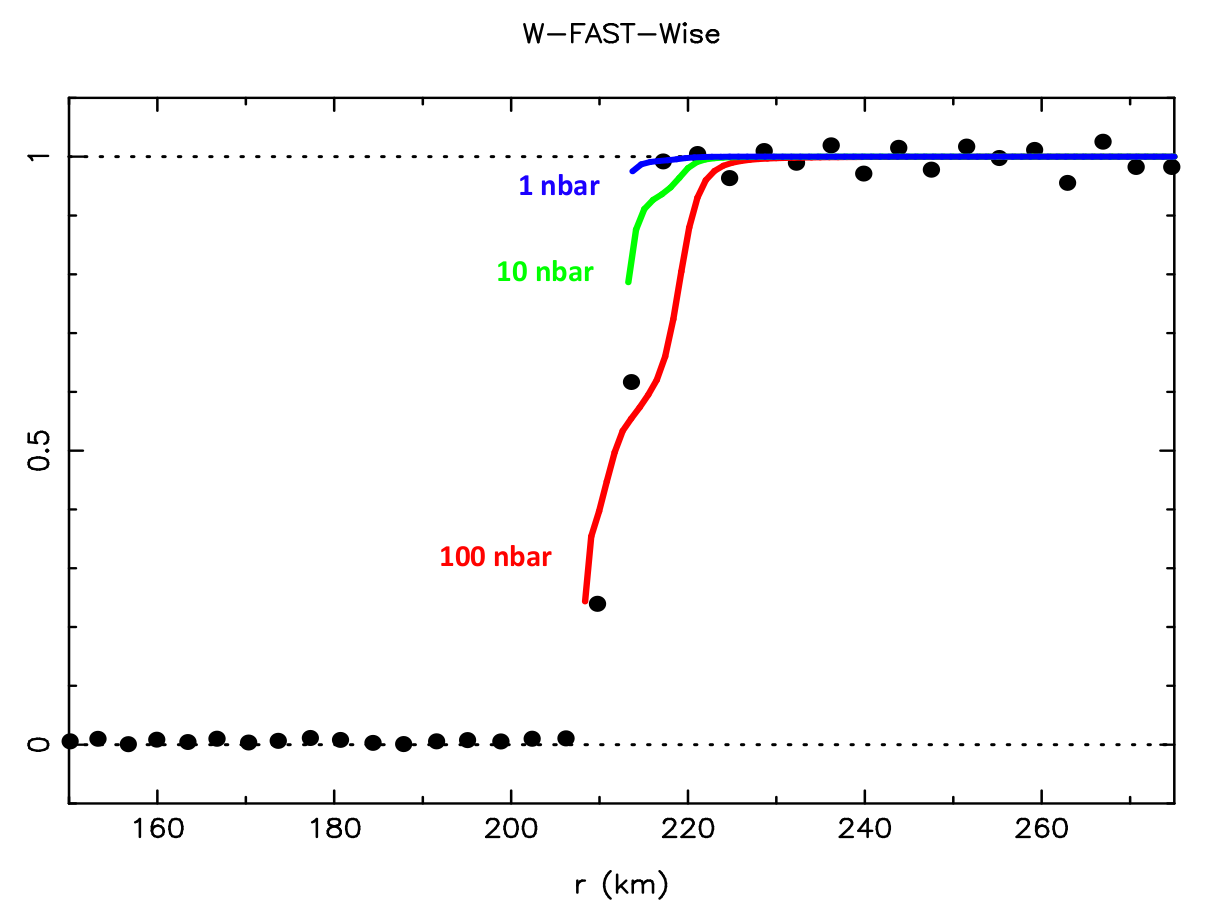}
    \caption{Comparison of various models assuming a Pluto-like atmosphere (see text for details) with the data obtained at the Wise station with the W-FAST instrument. The data points have been reprojected in the radial direction, merging the ingress and egress part of the light curve. The surface pressures used in each model are indicated next to each curve. The red model ($p_{\rm surf} = 100$ nbar) can clearly be discarded, considering its discrepancy with the observations. The green model ($p_{\rm surf} = 10$ nbar) could marginally accommodate the last point just outside the occultation and the rapid drop inside Huya's shadow. The blue model ($p_{\rm surf}= 1$ nbar) indicates that a 1~nbar atmosphere would go completely unnoticed in the data. We thus estimate that a conservative upper limit of $p_{\rm surf} = 10$ nbar of a Pluto-like atmosphere is provided by this light curve.}
    \label{atmos_models}
\end{figure}

A ray-tracing scheme \citep[see][]{Dias-Oliveira2015} then allows us to compare the W-FAST at Wise Observatory data (which is the best data set in terms of temporal sampling and flux dispersion) with models using various values of $p_{\rm surf}$. The results are displayed in  Fig. \ref{atmos_models}. It indicates that an upper limit of about $p_{\rm surf} = 10$ nbar can be placed for a Pluto-like Huya's atmosphere.

\section{Conclusions}
\label{conclu}

\begin{itemize}

\item This is the best ever stellar occultation produced by a TNO in terms of the number of chords, excluding Pluto and the August 8, 2020 occultation by the TNO 2002 MS$_4$ (Rommel et al. in prep.). Such a large number of chords, together with a relative small velocity of Huya with respect to the star as seen from Earth (8.07 km/s), has allowed to obtain the instantaneous limb of Huya with high precision.

\item An accurate area-equivalent diameter of 411.0 $\pm$ 7.3 km is determined from this stellar occultation. This diameter is larger than the radiometric area-equivalent diameter obtained for Huya using Herschel and Spitzer thermal data when the existence of the satellite is taken into account \citep[D = 406 $\pm$ 16 km; ][]{Fornasier2013}, but still compatible within error bars.

\item After a careful correction of the absolute magnitude of Huya by the rotational phase at the moment of the occultation and by the contribution of the satellite to the total flux of the system, an accurate geometric albedo at V-band is obtained:  $p_{\rm V}$ = 0.079 $\pm$ 0.004. This albedo is smaller than the radiometric albedo derived from Herschel and Spitzer measurements \citep[$p_{\rm V}$ = 0.083 $\pm$ 0.004; ][]{Fornasier2013}, but note that the latter is the albedo of the Huya's system, without subtracting the satellite contribution.

\item From the occultation results it is not clear what the three-dimensional shape of Huya is. The most likely shape is an oblate spheroid (Maclaurin), but a triaxial ellipsoid cannot be totally discarded. In any case, using the occultation results and the information on the rotational light curve (i.e., P$_{\rm preferred}$ = 6.725 h and $\Delta$m = 0.031 mag) we can constrain possible 3D solutions for Huya:

\begin{itemize}

\item If Huya's shape is a Maclaurin spheroid our preferred solution has axes: 2a = 435 km, 2b = 435 km, 2c = 233 km for an aspect angle of 32$^\circ$ and a density of $\sim$ 800 kg/m$^3$. The aspect angles for the other possible rotation periods (P $\sim$ 5.2 h and P $\sim$ 4.3 h) are smaller, which is very unlikely.

\item If Huya's shape is a Jacobi ellipsoid a family of solutions is possible: 2a = [440-455] km, 2b = [428-442] km, 2c = [253-261] km for aspect angles spanning from 31$^\circ$ to 43$^\circ$ and a minimum density of 859 kg/m$^3$. The minimum densities required for a Jacobi rotating with the other rotational periods are larger and very unlikely for a body of the Huya's size. 

\end{itemize}

\item If hydrostatic equilibrium for a homogeneous body is used to explain the occultation results and the rotational light curve, the densities appear to be too high (Jacobi solutions), or the aspect angle is too small (Maclaurin solutions). Alternative 3D shape solutions should invoke non-hydrostatic equilibrium shapes, which would not need so high densities.

\item We did not detect evidence of the Huya's satellite in any of the light curves (positive or negative) of this occultation.

\item No dense rings similar to the structures seen around the dwarf planet Haumea were detected orbiting Huya by means of this occultation, but very narrow ($\leq$ 0.1 km) and/or optically thin rings (opacity $\le$ 50\%) at different geometries cannot be totally discarded.

\item We obtained an upper limit of about $p_{\rm surf}$ = 10 nbar for a putative Pluto-like global atmosphere in Huya.

\end{itemize}


\begin{acknowledgements} 

P.S-S. acknowledges financial support by the Spanish grant AYA-RTI2018-098657-J-I00 ``LEO-SBNAF'' (MCIU/AEI/FEDER, UE). P.S-S., J.L.O., N.M., M.V-L. and R.D. acknowledge financial support from the State Agency for Research of the Spanish MCIU through the ``Center of Excellence Severo Ochoa'' award for the Instituto de Astrof\'isica de Andaluc\'ia (SEV-2017-0709), they also acknowledge the financial support by the Spanish grants AYA-2017-84637-R and PID2020-112789GB-I00, and the Proyectos de Excelencia de la Junta de Andaluc\'ia 2012-FQM1776 and PY20-01309. The research leading to these results has received funding from the European Union's Horizon 2020 Research and Innovation Programme, under Grant Agreement no. 687378, as part of the project ``Small Bodies Near and Far'' (SBNAF). Part of the research leading to these results has received funding from the European Research Council under the European Community's H2020 (2014-2020/ERC Grant Agreement no. 669416 ``LUCKY STAR''). Part of the work of M.P. was financed by a grant of the Romanian National Authority for Scientific Research and Innovation, CNCS - UEFISCDI, PN-III-P1-1.1-TE-2019-1504. This study was financed in part by the Coordena\c c\~ao de Aperfei\c coamento de Pessoal de N\'{i}vel Superior - Brasil (CAPES) - Finance Code 001 and the National Institute of Science and Technology of the e-Universe project (INCT do e-Universo, CNPq grant 465376/2014-2). The following authors acknowledge the respective CNPq grants: F.B-R 309578/2017-5; RV-M 304544/2017-5, 401903/2016-8; J.I.B.C. 308150/2016-3 and 305917/2019-6; M.A 427700/2018-3, 310683/2017-3 and 473002/2013-2; B.E.M. 150612/2020-6. G.B.R. thanks the support of CAPES and FAPERJ/PAPDRJ (E26/203.173/2016) grant. J.M.O. acknowledges financial support from the Portuguese Foundation for Science and Technology (FCT) and the European Social Fund (ESF) through the PhD grant SFRH/BD/131700/2017. E.F-V. acknowledges funding through the Preeminant Postdoctoral Program of the University of Central Florida. C.K., A.P. and R.S. have been supported by the grants K-125015 and K-138962 of the National Research, Development and Innovation Office (NKFIH, Hungary). E.P. acknowledges the Europlanet 2024 RI project funded by the European Union's Horizon 2020 Research and Innovation Programme (Grant agreement No. 871149). We are grateful to the CAHA and OSN staffs. This research is partially based on observations collected at the Centro Astron\'omico Hispano Alem\'an (CAHA) at Calar Alto, operated jointly by Junta de Andaluc\'ia and Consejo Superior de Investigaciones Cient\'ificas (IAA-CSIC). This research was also partially based on observation carried out at the Observatorio de Sierra Nevada (OSN) operated by Instituto de Astrof\'isica de Andaluc\'ia (CSIC). This article is also based on observations made in the Observatorios de Canarias del IAC with the Liverpool Telescope operated on the island of La Palma by the Instituto de Astrof\'isica de Canarias in the Observatorio del Roque de los Muchachos. Part of the results were based on observations taken at Pico dos Dias Observatory of the National Laboratory of Astrophysics (LNA/Brazil). Part of the data were collected during the photometric monitoring  observations with the robotic and remotely controlled observatory at  the University of Athens Observatory - UOAO \citep{Gazeas2016}. We thank the Adiyaman University Astrophysics Application and Research Center for their support in the acquisition of data with the ADYU60 telescope. This work has made use of data from the European Space Agency (ESA) mission {\it Gaia} (\url{https://www.cosmos.esa.int/gaia}), processed by the {\it Gaia} Data Processing and Analysis Consortium (DPAC, \url{https://www.cosmos.esa.int/web/gaia/dpac/consortium}). Funding for the DPAC has been provided by national institutions, in particular the institutions participating in the {\it Gaia} Multilateral Agreement. 

\end{acknowledgements}

%
%

\bibliographystyle{aa}


\bibliography{pablo_2020} 



\Online


\begin{appendix} 
\section{Additional tables}


\begin{table*}[!htb]
	\centering
	\caption{Observation details of the stellar occultation by the TNO Huya.}
	\label{ObservDetails}
	\resizebox{\textwidth}{!}{%
	\begin{tabular}{ccccccc} 
		\hline
\textbf{Chord} & \textbf{Observatory} 	& \textbf{Latitude (N)} & \textbf{Telescope diameter (m)}  & \textbf{Exposure time} & \textbf{Observer(s)} & \textbf{Detection}\\
\textbf{number} & \textbf{(Country)	}	& \textbf{Longitude (E)}  & \textbf{Detector} & \textbf{Cycle time}    & 					 & \textbf{$\sigma_{flux}$} \\
& \textbf{IAU code} 		 	& \textbf{Altitude (m)}  & 					   & \textbf{(seconds)} 	   & &\\
 \hline
 \hline
1 & TÜBITAK National Observatory-1  & 36$^\circ$ 49' 17.1"  & 1.0            & 2.0     & Y. Kilic, T. Ozisik    & Positive  \\
& (Turkey)                      & 30$^\circ$ 20' 08.0"  & SI 1100 Cryo   & 7.44    & O. Erece, S.E. Kilic   & 0.021    \\
& A84				              & 2470  & 			   &         & 		        & 			 \\	
 \hline
2 & TÜBITAK National Observatory-2  & 36$^\circ$ 49' 29.1" & 0.60             & 15.0     & Y. Kilic, T. Ozisik    & Positive  \\
& (Turkey)                      & 30$^\circ$ 20' 08.2" & FLI Proline 3041 & 20.0    & O. Erece, S.E. Kilic & 0.048  \\
& A84				              & 2455		  & 			   &         & 		        & 			 \\
\hline
3 & Romanian Academy, Astron. Obs. Cluj, Feleacu Station     & 46$^\circ$ 42' 37.6''	& 0.30              & 4.0    &     & Positive  \\
& (Romania)              & 23$^\circ$ 35' 35.7''    & SBIG STT-1603ME  & 4.5    & V. Turcu   & 0.038   \\
& ---			       & 783	                  & 			 &         &  & 			 \\	
 \hline
4 & ROASTERR-1 Observatory & 46$^\circ$ 49' 15.6''	& 0.30           & 0.2    &     & Positive  \\
& (Romania)              & 23$^\circ$ 35' 47.0''    & ASI 120MM      & 0.5    & L. Hudin   & 0.093  \\
& L04				       & 390	                  & 			 &         &  & 			 \\	
\hline
5 & W-FAST at Wise Observatory & 30$^\circ$ 35' 48.0''	  & 0.57        & 1.0     & 		 & Positive  \\
& (Israel)                   & 34$^\circ$ 45' 48.0''      & Andor Zyla	& 1.007 & G. Nir   & 0.015     \\
& 097				           & 900		  & 						& 		  & 		 & 			 \\
 \hline
6 & C18 at Wise Observatory           & 30$^\circ$ 35' 48.0''	  & 0.45        & 3.0     & 		 & Positive  \\
& (Israel)                   & 34$^\circ$ 45' 48.0''      & QSI 683 	& 4.46 & S. Kaspi   & 0.009 \\
& 097				           & 900		  & 						& 		  & 		 & 			 \\	
 \hline
7 & Astronomical Inst. of Romanian Academy-1 & 44$^\circ$ 24' 43.2"  & 0.50            & 2.0     & A. Sonka    & Positive  \\
& (Romania)                             &  26$^\circ$ 05' 38.2"   & FLI PL16803    & 4.2    & S. Anghel   & 0.047    \\
& 073				                      & 83		      & 			        &         &  & 			 \\	
 \hline
8 & Astronomical Inst. of Romanian Academy-2 & 44$^\circ$ 24' 48.0"	   & 0.50          & 1.0     &     & Positive  \\
& (Romania)             & 26$^\circ$ 05' 48.0"    & Andor Zyla    & 1.025    & D.A. Nedelcu   & 0.041    \\
& 073				      & 81		      & 			        &         &  & 			 \\	
 \hline
9 & ISTEK Belde Observatory       & 41$^\circ$ 01' 49.3''	  & 0.40                & 0.8     & M. Acar    & Positive  \\
& (Turkey)                      & 29$^\circ$ 02' 33.6''   & Sony IMX236LQJ CMOS & 0.8    & A. Ate\c{s}   & 0.088   \\
& ---			              & 150		      & 			        &         & C. Kayhan & 			 \\	
 \hline
10 & Amateur Observatory-1 & 44$^\circ$ 55' 05.0''	   & 0.15          & 4.0    &     & Positive  \\
& (Romania)           & 25$^\circ$ 58' 12.1''    & QHY6 CCD      & 6.0    & E. Petrescu   & 0.030 \\
& ---			    & 167		               & 			   &         &  & 			 \\	
\hline
11 & St. George Observatory & 45$^\circ$ 00' 25.2''	   & 0.18          & 3.0    &     & Positive  \\
& (Romania)              & 25$^\circ$ 58' 42.2''    & ATIK 460EX      & 4.0    & C. Danescu   & 0.026  \\
& L15				       & 242		              & 			   &         &  & 			 \\
\hline
12 & Amateur Observatory-2 & 44$^\circ$ 26' 36.5''	   & 0.20          & 0.20     &     & Positive  \\
& (Romania)             & 26$^\circ$ 31' 12.5''    & ASI 1600MM    & 0.25    & V. Dumitrescu   & 0.053 \\
& ---			      & 65		                & 			        &         &  & 			 \\	
 \hline
13 & Stardust Observatory   & 45$^\circ$ 38' 30.0''	& 0.20           & 8.0    &     & Positive  \\
& (Romania)              & 25$^\circ$ 37' 19.0''    & CCD Atik 383L+mono (KAF8300) & 12.0    & L. Curelaru   & 0.016 \\
& L13				       & 597	                  & 			   &         &  & 			 \\	
\hline
14 & Stardreams Observatory & 45$^\circ$ 12' 13.3''	  & 0.20           & 1.0    &     & Positive  \\
& (Romania)              & 26$^\circ$ 02' 44.2''    & ZWO ASI120MM-S & 1.01    & R. Gherase   & 0.074  \\
& L16				       & 379	                  & 			   &         &  & 			 \\	
\hline
15 & Martin S. Kraar Obs. / Weizmann Inst.  & 31$^\circ$ 54' 29.0''	  & 0.40        & 1.0     & 		 & Positive  \\
& (Israel)                               & 34$^\circ$ 48' 45.0''      & QHY 367C 	& 3.83    & I. Manulis   & 0.056   \\
& C78				                       & 107		  & 						& 		  & 		 & 			 \\	
\hline
$\star$ & Amateur Observatory-3     & 47$^\circ$ 05' 02.3''	& 0.40                  & 0.0333    &     & Positive  \\
& (Romania)              & 24$^\circ$ 23' 30.9''    & Watec 902H2 Ultimate  & 0.0333    & R. Truta   & 0.074  \\
& ---			       & 331	                  & 			 &         &  & 			 \\			       
\hline
16 & Amateur Observatory-4 & 44$^\circ$ 45' 41.5''	   & 0.20          & 0.20     &     & Positive  \\
& (Romania)             & 27$^\circ$ 20' 25.1''    & QHY 163M-CMOS    & 0.25    & D. Berte\c{s}teanu   & 0.093  \\
& ---			      & 41		                & 			        &         &  & 			 \\		
\hline
17 & Bacau Observatory  & 46$^\circ$ 33' 56.3''	& 0.35              & 2.0    &     & Positive  \\
& (Romania)              & 26$^\circ$ 54' 15.0''    & SBIG STL 6303E  & 9.0    & R. Anghel   & 0.026  \\
& L57			       & 170	                  & 			 &         &  & 			 \\	
\hline
18 & Galați Observatory-1   & 45$^\circ$ 25' 07.9''	  & 0.40             & 20.0    & J.O. Tercu & Positive  \\
& (Romania)              & 28$^\circ$ 01' 57.0''    & CCD SBIG STL-6303E & 24.8    & A.-M. Stoian   & 0.011  \\
& C73				       & 31	                  & 			   &         &  & 			 \\	
\hline
18 & Galați Observatory-2   & 45$^\circ$ 25' 07.9''	  & 0.20             & 20.0    & J.O. Tercu & Positive  \\
& (Romania)              & 28$^\circ$ 01' 57.0''    & CCD Atik 383L+Mono & 31.3    & A.-M. Stoian   & 0.031  \\
& C73				       & 31	                  & 			   &         &  & 			 \\	
\hline
19 & Barlad Observatory     & 46$^\circ$ 13' 54.1''	& 0.20           & 0.50    &     & Positive  \\
& (Romania)              & 27$^\circ$ 40' 10.2''    & ASI 1600      & 0.82    & C. Vantdevara   & 0.068  \\
& L22				       & 70	                  & 			 &         &  & 			 \\	
\hline
\hline

& QOS Observatory  		& 48$^\circ$ 50' 54.0''  & 0.30	        & 0.2 	     & T. O. Dementiev	 & Negative \\
& (Ukraine)				& 26$^\circ$ 43' 12.0''  & ZWO ASI 174MM  & 0.200016   & O. M. Kozhukhov   & 0.106    \\   							    					
& L18					& 352				   & 			    & 		     & 			         & 			\\
\hline
& \c{C}ukurova University	& 37$^\circ$ 03' 23.0''  & 0.50	             & 90 	     & A. Solmaz	 & Negative \\
& (Turkey)				    & 35$^\circ$ 21' 04.0''  & Orion Parsec 8300M  & 108      & M. Tekes   &  0.027   \\   							    				
& ---					    & 130				   & 			    & 		        & 			         & 			\\
		\hline
	\end{tabular}}
	{\tiny
This Table includes the sites from where the occultation was detected and the closest negatives used to constrain the instantaneous shape of the object. $\sigma_{flux}$ is the flux dispersion, i.e. the standard deviation of the normalized flux (outside of the occultation in the case of positive detections). Sites are sorted by their distance to the center of the predicted shadow path, from West to East. The data from the site marked with $\star$ was not used in the final analysis of the occultation (see Section \ref{analysis} for a detailed explanation).}	
\end{table*}


\onecolumn

\label{append1}
\begin{longtable}{ccccc}

	\caption{Summary table of the observing campaign.}\\

		\hline
\textbf{Observatory} 	& \textbf{Latitude (N)} & \textbf{Telescope}  & \textbf{Observer(s)} & \textbf{Observation}\\
\textbf{(Country)	}	& \textbf{Longitude (E)}  & \textbf{aperture} &    & 	\\
\textbf{IAU code}	& \textbf{Altitude (m)}  & 	\textbf{(m)}				   &  	   & \\

\hline
\endfirsthead  

\multicolumn{5}{c}{\tablename\ \thetable\ - Summary table of the observing campaign \textit{(continued from previous page)}} \\	
	
\hline
\textbf{Observatory} 	& \textbf{Latitude (N)} & \textbf{Telescope}  & \textbf{Observer(s)} & \textbf{Observation}\\
\textbf{(Country)	}	& \textbf{Longitude (E)}  & \textbf{aperture} &    & 	\\
\textbf{IAU code}	& \textbf{Altitude (m)}  & 	\textbf{(m)}				   &  	   & \\ 
\hline
\endhead
\hline \multicolumn{4}{c}{\textit{continued on next page}} \\
\endfoot
\hline
\endlastfoot 

	\label{SummaryObserv}
 
Sant Esteve Sesrovires 		& 41$^\circ$ 29' 37.5''  & 0.40  & C. Schnabel    &	Negative \\
(Spain)				        & 01$^\circ$ 52' 21.1''  &       &                & \\   							    					
--							& 180					  & 			&				& 	\\	
\hline
Sabadell Observatory  		& 41$^\circ$ 33' 00.2''  & 0.50  & C. Perello    &	Negative \\
(Spain)				        & 02$^\circ$ 05' 24.6''  &       & A. Selva      & \\   							    					
619		            & 224					  & 			&				& 	\\
\hline
Normandy           & 49$^\circ$ 36' 26.4''  & 0.14   &  J. Lecacheux   &	Negative \\
(France)		   & 358$^\circ$ 46' 03.3''  &        &                  &      \\   							    					
--				   & 12      				  & 							& 		     & 		\\
\hline	
Northolt Branch Observatories       &  51$^\circ$ 33' 16.8"  &  0.25  & G. Wells       &	Negative \\
(UK)		                        &  359$^\circ$ 37' 41.1"  &       & D. Bamberger  &      \\   							    					
Z80              &  55     				  & 							& 		     & 		\\
\hline
Almalex Observatory       &  53$^\circ$ 50' 15.4"  &  0.28  & A. Pratt    &	Negative \\
(UK)		                &  358$^\circ$ 23' 32.0"  &        &                  &      \\   							    					
Z92				            &  114     				  & 							& 		     & 		\\
\hline	
NOAK Observatory   & 39$^\circ$ 39' 08.8''  & 0.25   &  N. Sioulas   &	Bad \\
(Greece)		   & 20$^\circ$ 48' 59.7''  &        &                  &  weather    \\   							    					
L02				   & 546					  & 							& 		     & 		\\
\hline
Univ. of Athens Observatory - UOAO  & 37$^\circ$ 58' 06.8''  & 0.40   &  K. Gazeas   &	Negative \\
(Greece)				        & 23$^\circ$ 47' 00.1''  &        & E. Karampotsiou &      \\   							    					
--							    & 250					  & 							& 		     & 		\\
\hline	
Ellinogermaniki Agogi Observatory   & 37$^\circ$ 59' 52.3''  & 0.40   &  V. Tsamis   &	Negative \\
(Greece)				            & 23$^\circ$ 53' 36.1''  &        & K. Tigani &      \\   							    					
C68							        & 162					  & 							& 		     & 		\\
\hline
Konkoly Observatory Budapest & 47$^\circ$ 29' 59.3''  & 0.60  &  A. Pal   &	Bad \\
(Hungary)				     & 18$^\circ$ 57' 51.1''  &  & R. Szakats &  weather \\   							    					
053							 & 469					  & 							& 		     & 		\\
\hline	
Ulupınar Observatory           & 40$^\circ$ 06' 01.0''	& 0.40      & C. Puskullu & Negative  \\
(Turkey)                       & 26$^\circ$ 28' 32.0''  &           &            &             \\
--        			           & 410	                & 			&             & 			 \\						    
\hline
Berlin                 &  52$^\circ$ 30' 58.0"  &  0.20  & C. Weber       &	Negative \\
(Germany)	           &  13$^\circ$ 25' 40.0"  &       &                 &      \\   							    					
--				       &  37     				  & 							& 		     & 		\\
\hline
Konkoly Observatory Piszkestető & 47$^\circ$ 55' 06.0''  & 1.00   &  A. Pal   &	Negative \\
(Hungary)				        & 19$^\circ$ 53' 41.7''  &          & R. Szakats &      \\   							    					
561							    & 944					  & 							& 		     & 		\\
\hline
TÜBITAK National Observatory-1  & 36$^\circ$ 49' 17.1"  & 1.00      & Y. Kilic, T. Ozisik    & Positive  \\
(Turkey)                        & 30$^\circ$ 20' 08.0"  &          &  O. Erece, S.E. Kilic  &     \\
A84				                & 2472		            & 			   &  & \\	
 \hline
TÜBITAK National Observatory-2  & 36$^\circ$ 49' 29.1" & 0.60    & Y. Kilic, T. Ozisik    & Positive  \\
(Turkey)                        & 30$^\circ$ 20' 08.2" &         & O. Erece, S.E. Kilic   &          \\
A84				                & 2472		  & 			   &         & 		    			 \\
\hline
Akdeniz University Observatory & 36$^\circ$ 49' 27.0''	& 0.25      & V. Bak{\i}\c{s}  & Technical  \\
(Turkey)                       & 30$^\circ$ 20' 08.0''   &           & H. Bak{\i}\c{s}   &  problems   \\
--        			           & 2490	                & 			& Z. Eker     & 			 \\	
\hline	
Mt. Suhora Observatory &  49$^\circ$ 34' 09.0''     &  0.60  & W. Ogłoza    &  Bad \\
(Poland)               &  20$^\circ$ 04' 03.0''     &        &              &   weather \\    
-- 					   & 1009   				    & 		 & 		        & 			\\
\hline 					   
Borowiec Observatory  	& 52$^\circ$ 16' 37.2''  & 0.40  &  A. Marciniak  &	Bad \\
(Poland)				& 17$^\circ$ 04' 28.6''  &         &                & weather\\   							    					
187					    & 123					 & 		   & 		        &       \\
\hline 	
Romanian Acad., Astron. Obs. Cluj, Feleacu Station   & 46$^\circ$ 42' 37.6''	    & 0.30      &            & Positive  \\
(Romania)               & 23$^\circ$ 35' 35.7''     &           & V. Turcu   &            \\
--                      & 783	                    & 			 &           & 			   \\	
 \hline
ROASTERR-1 Observatory & 46$^\circ$ 49' 15.6''	  & 0.30        &            & Positive  \\
(Romania)              & 23$^\circ$ 35' 47.0''    &             & L. Hudin   &           \\
L04				       & 390	                  & 			 &         &  	 \\	
\hline
W-FAST at Wise Observatory & 30$^\circ$ 35' 48.0''  & 0.57   & 		    & Positive  \\
(Israel)                   & 34$^\circ$ 45' 48.0''  &        & G. Nir   &            \\
097				           & 900		  & 						& 		  & 		  \\
\hline
C18 at Wise Observatory   & 30$^\circ$ 35' 48.0''	& 0.45   & 		      & Positive  \\
(Israel)           & 34$^\circ$ 45' 48.0''  &        & S. Kaspi   &             \\
097				   & 900		  & 						& 		  & 		  \\	
 \hline
Astronomical Inst. of Romanian Academy-1  & 44$^\circ$ 24' 43.2"  & 0.50      & A. Sonka    & Positive  \\
(Romania)                               &  26$^\circ$ 05' 38.2" &           & S. Anghel   &          \\
073				                        & 83		                &        &         & 		 \\	
\hline
Astronomical Inst. of Romanian Academy-2 & 44$^\circ$ 24' 48.0"    & 0.50          &                & Positive  \\
(Romania)             & 26$^\circ$ 05' 48.0"    &           & D.A. Nedelcu   &          \\
073				      & 81		      & 		&         &  			 \\	
 \hline
Astroclubul Bucure\c{s}ti   & 44$^\circ$ 24' 49.0''      &  0.40  & Z. Deak        &  Technical \\
 (Romania)              & 26$^\circ$ 05' 48.1''      &        &                &   problems \\
-- 					    & 69   				   & 			    & 		        & 			\\
\hline
ISTEK Belde Observatory       & 41$^\circ$ 01' 49.3''	& 0.40      & M. Acar       & Positive  \\
(Turkey)                      & 29$^\circ$ 02' 33.6''   &           & A. Ate\c{s}   &            \\
--        			          & 150		                & 			& C. Kayhan     & 			 \\	
 \hline
ISTEK Belde Observatory       & 41$^\circ$ 01' 49.3''	& 0.35      & M. Acar       & Technical  \\
(Turkey)                      & 29$^\circ$ 02' 33.6''   &           & A. Ate\c{s}   &  problems   \\
--        			          & 150		                & 			& C. Kayhan     & 			 \\
\hline 
ISON-Uzhgorod Observatory ``Derenivka'' & 48$^\circ$ 33' 48.2''    & 0.40  &  V. Kudak  &	Bad \\
(Ukraine)				& 22$^\circ$ 27' 12.6''   &         &    V. Perig  & weather \\   							    					
K99					    & 213					 & 		   & 		        &       \\
\hline
Amateur Observatory-1      & 44$^\circ$ 55' 05.0''	   & 0.15       &               & Positive  \\
(Romania)                       & 25$^\circ$ 58' 12.1''    &            & E. Petrescu   &            \\
--            			        & 167		               & 			&         &   			        \\	
\hline
St. George Observatory & 45$^\circ$ 00' 25.2''	  & 0.18    &               & Positive  \\
(Romania)              & 25$^\circ$ 58' 42.2''    &         & C. Danescu    &           \\
L15				       & 242		              & 		&               & 			 \\
\hline
Amateur Observatory-2  & 44$^\circ$ 26' 36.5''	   & 0.20       &                   & Positive  \\
(Romania)                 & 26$^\circ$ 31' 12.5''  &            & V. Dumitrescu     &            \\
--        			      & 65		               & 		    &                   & 			 \\	
 \hline
Stardust Observatory   & 45$^\circ$ 38' 30.0''	& 0.20      &               & Positive  \\
(Romania)              & 25$^\circ$ 37' 19.0''  &           & L. Curelaru   &  \\
L13				       & 597	                &         &  & 			 \\	
\hline
Stardreams Observatory & 45$^\circ$ 12' 13.3''	  & 0.20       &            & Positive  \\
(Romania)              & 26$^\circ$ 02' 44.2''    &            & R. Gherase   &   \\
L16				       & 379	                  & 			   &         &  		 \\	
\hline
Martin S. Kraar Obs./ Weizmann Inst.   & 31$^\circ$ 54' 29.0''	  & 0.40       & 		 & Positive  \\
(Israel)                               & 34$^\circ$ 48' 45.0''    &             & I. Manulis   &    \\
C78				                       & 107		  & 						& 		  & 	 \\	
\hline
Amateur Observatory-3     & 47$^\circ$ 05' 02.3''	& 0.40  &           & Positive  \\
(Romania)                 & 24$^\circ$ 23' 30.9''   &       & R. Truta   &   \\
--                        & 331	                  & 			 &         &  	 \\			       
\hline
Amateur Observatory-4   & 44$^\circ$ 45' 41.5''	   & 0.20    &     & Positive  \\
(Romania)               & 27$^\circ$ 20' 25.1''    &         & D. Berte\c{s}teanu  &  \\
--                	    & 41		               & 			        &         &  	 \\		
\hline
Bacau Observatory     & 46$^\circ$ 33' 56.3''	& 0.35        &            & Positive  \\
(Romania)                 & 26$^\circ$ 54' 15.0''    &           & R. Anghel   &   \\
L57                        & 170	                  & 			 &         &    	 \\	
\hline
Galați Observatory-1        & 45$^\circ$ 25' 07.9''	  & 0.40            & J.O. Tercu & Positive  \\
(Romania)                   & 28$^\circ$ 01' 57.0''   &                 & A.-M. Stoian   &   \\
C73				            & 31	                  & 			    &  & 			 \\	
\hline
Galați Observatory-2   & 45$^\circ$ 25' 07.9''	  & 0.20         & J.O. Tercu     & Positive  \\
(Romania)                   & 28$^\circ$ 01' 57.0''   &              & A.-M. Stoian  &  \\
C73				            & 31	                  & 			   &         &  	 \\	
\hline
Barlad Observatory          & 46$^\circ$ 13' 54.1''	  & 0.20    &     & Positive  \\
(Romania)                   & 27$^\circ$ 40' 10.2''   &         & C. Vantdevara   &  \\
L22                          & 70	                  & 		&       & 			 \\
\hline
QOS Observatory  		& 48$^\circ$ 50' 54.0''  & 0.30	 & T. O. Dementiev	 & Negative \\
(Ukraine)				& 26$^\circ$ 43' 12.0''  &        & O. M. Kozhukhov   &    \\   							    					
L18					    & 352				   & 			    & 		     & 		\\
\hline
\c{C}ukurova University	& 37$^\circ$ 03' 23.0''  & 0.50	     & A. Solmaz	 & Negative \\
(Turkey)				& 35$^\circ$ 21' 04.0''  &           & M. Tekes   &    \\   							    				
--					    & 130				   & 			    & 		        & 			\\
\hline 
Odessa Astronomical Observatory, Mayaky   & 46$^\circ$ 23' 49.1''  & 0.80	     & V. Kashuba	 & Negative \\
(Ukraine)				        & 30$^\circ$ 16' 16.6''  &           & V. Zhukov   &    \\   							    				
583					            & 18				     & 			 & 		        & 			\\
\hline
Adiyaman      	        & 38$^\circ$ 13' 31.2''    & 0.60  &  M.  Żejmo  &	Bad \\
(Turkey)				& 37$^\circ$ 45' 06.1''   &         &                   & weather \\   							    					
--					    & 690					 & 		   & 		            &       \\
\hline 
Ondokuz Mayıs University       & 41$^\circ$ 22' 04.0''	& 0.35      & S. Kalkan & Technical  \\
(Turkey)                       & 36$^\circ$ 12' 06.0''  &           &            & problems    \\
--        			           & 151	                & 			&             & 			 \\	
\hline
Moletai               	& 55$^\circ$ 18' 57.5"   & 0.35  &  E. Pakštienė  &	Bad \\
(Lithuania)				& 25$^\circ$ 33' 48.0"   &         &                & weather\\   							    					
--					    & 200					 & 		   & 		        &       \\
\hline 	
İnönü University Observatory  & 38$^\circ$ 19' 24.0''	& 0.35      & T. Ozdemir  & Bad  \\
(Turkey)                       & 38$^\circ$ 26' 13.0''   &           &            &  weather   \\
--        			           & 130	                & 			&             & 			 \\	
\hline
Makes Observatory-La Reunion    & -21$^\circ$ 11' 56.1''  &  0.18     &  J.P. Teng   &	    Negative \\
(France)		        &  55$^\circ$ 24' 36.3''  &        &                  &      \\   							    					
--				        &  997     				  & 							& 		     & 		\\
\hline	
Sainte Marie-La Reunion    &  -20$^\circ$ 53' 48.5"  &  0.20  & B. Mondon    &	Negative \\
(France)		            &  55$^\circ$ 34' 00.1"  &        &                  &      \\   							    					
--				            &  54     				  & 							& 		     & 		\\
\hline
Ataturk University      & 39$^\circ$ 54' 17.1''    & 0.50  &  C. Yesilyaprak   &	Negative \\
(Turkey)				& 41$^\circ$ 14' 40.1''   &         &   O. Satir, M.S. Niaei &           \\   							    					
--					    & 1857					 & 		   & 	E. Atalay	            &       \\
\hline
\end{longtable}
This table includes all the telescopes and observers that supported the Huya occultation campaign. Sites are sorted by their distance to the
center of the predicted shadow path, from West to East.


\begin{table*}[!htb]
	\centering
	\caption{Ingress and egress times derived from the Huya occultation light curves.}
	\label{Chords}
	\resizebox{\textwidth}{!}{%
	\begin{tabular}{clccc}   
		\hline	
		\hline	    
\textbf{\#chord} & \textbf{Observatory, Country}  &  \textbf{Ingress (UT)} & \textbf{Egress (UT)} & \textbf{Chord size (km)} \\
\hline
1 &TÜBITAK National Observatory-1, Turkey  &	00:52:34.387$\pm$2.592 &  00:53:00.703$\pm$0.142 & 212.4$\pm$22.1 \\
\hline
2 &TÜBITAK National Observatory-2, Turkey  &	00:52:30.407$\pm$5.500 &  00:52:52.439$\pm$5.500 & 177.8$\pm$88.8 \\
 \hline
3 &Romanian Acad., Astron. Obs. Cluj, Feleacu Sta., Romania  &	00:53:51.515$\pm$0.115 &  00:54:37.293$\pm$0.089 & 369.4$\pm$1.7 \\
\hline
4 &ROASTERR-1 Observatory, Romania  &	No data &  00:54:38.748$\pm$0.030 & --- \\
\hline
5 &W-FAST-Wise Observatory, Israel  &	00:51:21.416$\pm$0.016 &  00:52:07.918$\pm$0.023 & 375.3$\pm$0.3 \\
 \hline
6 &C18 at Wise Observatory, Israel  &	00:51:21.110$\pm$0.040 &  00:52:07.851$\pm$0.040 & 377.2$\pm$0.7 \\
 \hline
7 &Astronomical Inst. Romanian Academy-1, Romania  &	00:53:30.014$\pm$0.998 &  00:54:18.863$\pm$0.098 & 394.2$\pm$8.8 \\
 \hline 
8 &Astronomical Inst. Romanian Academy-2, Romania  &	00:53:31.111$\pm$0.041 &  00:54:18.814$\pm$0.041 & 385.0$\pm$0.7 \\
\hline 
9 &ISTEK Belde Observatory, Turkey  &	00:52:59.908$\pm$0.062 &  00:53:50.451$\pm$0.083 & 407.9$\pm$1.2 \\
\hline
10 &Amateur Observatory-1, Romania  &	00:53:33.562$\pm$0.444 &  00:54:23.630$\pm$0.174 & 404.1$\pm$5.0 \\
\hline
11 &St. George Observatory, Romania  &	00:53:34.446$\pm$0.083 &  00:54:24.868$\pm$0.077 & 406.9$\pm$1.3 \\
\hline
12 &Amateur Observatory-2, Romania  &	00:53:30.484$\pm$0.020 &  00:54:21.217$\pm$0.200 & 409.4$\pm$1.8 \\
\hline
13 &Stardust Observatory, Romania  &	00:53:38.100$\pm$0.150 &  00:54:29.800$\pm$1.850 & 417.2$\pm$16.1 \\
\hline
14 &Stardreams Observatory, Romania  &	00:53:37.111$\pm$0.098 &  00:54:29.122$\pm$0.099 & 419.7$\pm$1.6 \\
\hline
15 &Martin S. Kraar Obs., Israel  &	00:51:30.046$\pm$0.100 &  00:52:23.578$\pm$1.148 & 432.0$\pm$10.1 \\
\hline
$\star$ &Amateur Observatory-3, Romania  &	00:00:00.000$\pm$0.007 &  00:00:43.218$\pm$0.007 & 348.8$\pm$0.1 \\
	\hline
16 &Amateur Observatory-4, Romania  &	00:53:33.840$\pm$0.027 &  00:54:25.060$\pm$0.027 & 413.4$\pm$0.4 \\
\hline
17 & Bacau Observatory, Romania  & 00:53:46.520$\pm$3.400 &  00:54:33.919$\pm$0.083 & 382.5$\pm$28.1 \\
\hline
18 &Galați Observatory-1 and 2, Romania  &	00:53:40.500$\pm$0.440 &  00:54:25.420$\pm$0.440 & 362.5$\pm$7.1 \\
\hline
19 &Barlad Observatory, Romania  &	00:53:51.192$\pm$0.047 &  00:54:31.753$\pm$0.073 & 327.3$\pm$1.0 \\
\hline
	\end{tabular}}
Sites are sorted by their distance to the center of the predicted shadow path, from West to East. The chord marked with $\star$ was not used in the final analysis of the occultation (see Section \ref{analysis}).
\end{table*}


\begin{table}
	\centering
	\caption{Parameters of the best-fitting ellipse.}
	\label{limbfit}
	{%
    \renewcommand{\arraystretch}{1.35}
	\begin{tabular}{c|c}   
(a$^\prime$ , b$^\prime$) & (217.6 $\pm$ 3.5 km , 194.1 $\pm$ 6.1 km) \\
\hline
($f_{\rm c}$ , $g_{\rm c}$) & (2984.7 $\pm$ 3.2 km , -1850.9 $\pm$ 1.7 km) \\
\hline
P$^\prime$ & 55.2$^\circ \pm$ 9.1 \\
\hline
$D_{\rm eq}$ & 411.0 $\pm$ 7.3 km \\
\hline
$p_{\rm V}$ & 0.079 $\pm$ 0.004 \\
	\end{tabular}}

Parameters of the best-fitting ellipse obtained from the occultation chords plus the closest non-detection constraint (Figure \ref{elliptical_fit}). The area-equivalent diameter ($D_{\rm eq}$ ) and geometric albedo ($p_{\rm V}$) derived from this limb fit are also included (see Section \ref{shape-albedo} for details).
\end{table}


\begin{table*}[!hpbt]
	\centering
	\caption{Ring constraints not accounting for dead-times.}
	\label{ringconstraints}
	\resizebox{\textwidth}{!}{%
	\begin{tabular}{c|cccc}   
		\hline	    
   \textbf{Observatory} & \textbf{w} & \textbf{op$_{\rm min}$} & \textbf{w(op=50\%)} & \textbf{w(op=100\%)}\\
   \textbf{(Telesc., Country)} & \textbf{(km)} & & \textbf{(km)} & \textbf{(km)}\\
\hline
\hline
TÜBITAK National Observatory-1 & 16.1 & 6.3\% & 2.0 & 1.0\\ 
(1.00 m, Turkey) & & & &\\
\hline
TÜBITAK National Observatory-2 & 121.1 & 14.4\% & 34.9 & 17.4\\ 
(0.60 m, Turkey) & & & &\\
\hline
Romanian Academy, Astronomical Observatory Cluj, Feleacu Station & 32.3 & 11.4\% & 7.4 & 3.7\\ 
(0.30 m, Romania) & & & &\\
\hline
ROASTERR-1 Observatory & 1.6 & 27.9\% & 0.9 & 0.5\\ 
(0.30 m, Romania) & & & &\\
\hline
W-FAST at Wise Observatory & 8.1 & 4.5\% & 0.7 & 0.4\\
(0.57 m, Israel) & & & &\\
\hline
C18 at Wise Observatory  & 24.2 & 2.7\% & 1.3 & 0.7\\
(0.45 m, Israel) & & & &\\
\hline
Astronomical Inst. of Romanian Academy-1 & 16.1 & 14.1\% & 4.6 & 2.3\\ 
(0.50 m, Romania) & & & & \\
\hline
Astronomical Inst. of Romanian Academy-2 & 8.1 & 12.3\% & 2.0 & 1.0\\ 
(0.50 m, Romania) & & & &\\
\hline
ISTEK Belde Observatory & 6.5 & 26.4\% & 3.4 & 1.7\\ 
(0.40 m, Turkey) & & & &\\
\hline
Amateur Observatory-1 & 32.3 & 9.0\% & 5.8 & 2.9\\ 
(0.15 m, Romania) & & & & \\
\hline
St. George Observatory & 24.2 & 7.8\% & 3.8 & 1.9\\ 
(0.18 m, Romania) & & & & \\
\hline
Amateur Observatory-2 & 1.6 & 15.9\% & 0.5 & 0.3\\ 
(0.20 m, Romania) & & & &\\
\hline
Stardust Observatory & 64.6 & 4.8\% & 6.2 & 3.1\\ 
(0.20 m, Romania) & & & &\\
\hline
Stardreams Observatory & 8.1 & 22.2\% & 3.6 & 1.8\\ 
(0.20 m, Romania) & & & &\\
\hline
Martin S. Kraar Obs./Weizmann Inst. & 8.1 & 16.8\% & 2.7 & 1.4\\ 
(0.40 m, Israel) & & & &\\
\hline
Amateur Observatory-3 & 0.3 & 22.2\% & 0.1 & 0.1\\ 
(0.40 m, Romania) & & & &\\
\hline
Amateur Observatory-4 & 1.6 & 27.9\% & 0.9 & 0.5\\ 
(0.20 m, Romania) & & & &\\
\hline
Bacau Observatory & 16.1 & 7.8\% & 2.5 & 1.3\\ 
(0.35 m, Romania) & & & &\\
\hline
Galați Observatory-1 & 161.4 & 3.3\% & 10.7 & 5.3\\ 
(0.40 m, Romania) & & & & \\
\hline
Galați Observatory-2 & 161.4 & 9.3\% & 30.0 & 15.0\\ 
(0.20 m, Romania) & & & &\\
\hline
Barlad Observatory & 4.0 & 20.4\% & 1.6 & 0.8\\ 
(0.20 m, Romania) & & & &\\
\hline
QOS Observatory (NEGATIVE) & 1.6 & 31.8\% & 1.0 & 0.5\\ 
(0.30 m, Ukraine) & & & &\\
\hline
	\end{tabular}}
	Ring width (w) and minimum opacity (op$_{\rm min}$) for a 3$\sigma$ detection, obtained from all the telescopes that detected the occultation and the closest negative one. In the third and fourth column we estimate the ring width detectable at 3$\sigma$ level for opacities of 50\% and 100\% (completely opaque ring), respectively.
\end{table*}

\end{appendix}


\end{document}